\documentclass{ws-ijmpd}
\usepackage[section]{placeins}
\begin{document}

\markboth{K. Lin, F.-H. Ho, and W.-L. Qian}
{Charged Einstein-\ae ther black holes in $n$-dimensional spacetime}

 \newcommand{\bq}{\begin{equation}}
 \newcommand{\eq}{\end{equation}}
 \newcommand{\bqn}{\begin{eqnarray}}
 \newcommand{\eqn}{\end{eqnarray}}
 \newcommand{\nb}{\nonumber}
 \newcommand{\lb}{\label}
\newcommand{\PRL}{Phys. Rev. Lett.}
\newcommand{\PL}{Phys. Lett.}
\newcommand{\PR}{Phys. Rev.}
\newcommand{\CQG}{Class. Quantum Grav.}
\newcommand{\hong}[1]{\textcolor{red}{#1}}

\title{Charged Einstein-\ae ther black holes in $n$-dimensional spacetime}

\author{Kai Lin}
\address{Hubei Subsurface Multi-scale Imaging Key Laboratory, Institute of Geophysics and Geomatics, China University of Geosciences, 430074, Wuhan, Hubei, China\\
Escola de Engenharia de Lorena, Universidade de S\~ao Paulo, CEP 12602-810, Lorena, SP, Brazil\\
Instituto de F\'isica e Qu\'imica, Universidade Federal de Itajub\'a, CEP 37500-903, Itajub\'a, Brazil\\
lk314159@hotmail.com}

\author{Fei-Hung Ho}
\address{School of Science, Jimei University, 361021, Xiamen, Fujian, China\\
Institute  for Advanced Physics $\&$ Mathematics, Zhejiang University of Technology, Hangzhou 310023, China\\
fhho@jmu.edu.cn}

\author{Wei-Liang Qian}
\address{Escola de Engenharia de Lorena, Universidade de S\~ao Paulo, CEP 12602-810, Lorena, SP, Brazil\\
Faculdade de Engenharia de Guaratinguet\'a, Universidade Estadual Paulista, 12516-410, Guaratinguet\'a, SP, Brazil\\
School of Physical Science and Technology, Yangzhou University, 225002, Yangzhou, Jiangsu, China\\
wlqian@usp.br}

\maketitle

\begin{history}
\received{Day Month Year}
\revised{Day Month Year}
\end{history}

\begin{abstract}

In this work, we investigate the $n$-dimensional charged static black hole solutions in the Einstein-\ae ther theory.
By taking the metric parameter $k$ to be $1,0$, and $-1$, we obtain the spherical, planar, and hyperbolic spacetimes respectively.
Three choices of the cosmological constant, $\Lambda>0$, $\Lambda=0$ and $\Lambda<0$, are investigated, which correspond to asymptotically de Sitter, flat and anti-de Sitter spacetimes.
The obtained results show the existence of the universal horizon in higher dimensional cases which may trap any particle with arbitrarily large velocity.
We analyze the horizon and the surface gravity of 4- and 5-dimensional black holes, and the relations between the above quantities and the electrical charge.
It is shown that when the aether coefficient $c_{13}$ or the charge $Q$ increases, the outer Killing horizon shrinks and approaches the universal horizon.
Furthermore, the surface gravity decreases and approaches zero in the limit $c_{13}\rightarrow\infty$ or $Q\rightarrow Q_e$, where $Q_e$ is the extreme charge.
The main features of the horizon and surface gravity are found to be similar to those in $n=3$ case, but subtle differences are also observed.

\end{abstract}

\keywords{Einstein-\ae ther gravity; higher dimensional black hole; universal horizon.}

\ccode{04.50.Kd, 04.20.Jb, 04.70.Dy}


\section{Introduction}

As one of the cornerstones of modern physics, the principle of the invariance of the speed of light entails Lorentz symmetry.
Experimentally, superluminal particles have not been found, which strongly indicates that Lorentz symmetry is well reserved in the lower energy region.
Nevertheless, in the developments of theoretical physics and astrophysics, more and more models and theories require the breaking of Lorentz invariance (LI) in the higher energy region.
For example, LI breakdown has been introduced in the context of the beta decay theory of Fermi\cite{Fermi}, noncommutative field theories\cite{NCF1,NCF2} and string theories when the perturbative string vacuum is unstable\cite{stringLIV}.
Furthermore, the most notorious difficulty in general relativity is related to its quantum mechanical corrections at the Planck scale.
In fact, the quantum cosmology and the details of the very early universe, for example, the details of inflation of very early universe, are still mostly unknown to us.
As one of the possibilities, Ho\v{r}ava-Lifshitz (HL) gravity\cite{Horava} has been introduced as an alternative to another rather large framework, the string theory.
The price one has to pay is that, in the ultraviolet limit, LI is violated.
Besides, some models of dark matter also require the violation of LI\cite{Mukohyama}.

HL gravity was proposed in 2009\cite{Horava}.
In HL theory, the LI is broken, the time derivative terms are of the second order at most, while the spatial derivative terms are up to 6th order, and therefore the renormalizability is achieved.
However, the violation of LI only appears in the higher energy region, in the sense that the higher order spatial derivative terms are insignificant in the lower-energy region, thus the HL theory is not in contradiction with today's physical experiments.
Since the advent of HL gravity, the theory is plagued by some difficulties, such as strong coupling, instability and ghost problems.
So some new physical conditions, such as the detailed balance condition and new scalar fields $A$ and $\varphi$ with $U(1)$ symmetry, are introduced to handle the above difficulties.
In our recent works, we studied several versions of HL gravity which in principle avoid all the above difficulties\cite{HC,newversion1,newversion2,newversion3,newversion4,newversion5}, and it is shown that our model of HL passes the test of Post-Newtonian experiments\cite{PPN1,PPN2,PPN3}.
It's worth noting that the work\cite{Mukohyama} indicate that the constant term of projectable HL theory might be a dark matter candidate.

Recently, it is shown that the HL theory becomes an effective Einstein-\ae ther theory at the lower energy region\cite{BPS1,BPS2}, and this conclusion may greatly simplify the study of the HL gravity.
Einstein-\ae ther theory also violates the LI, where superluminal particles are allowed.
Therefore, these modified gravitational theories face a question in black hole physics: Do black holes still exist in those scenarios without LI?
According to the modified gravity without LI, the killing horizon cannot be used to define a black hole because superluminal particles inside of killing event horizon could still travel through the surface, and therefore escape from inside the killing horizon.
So, does it mean that the singularity will be exposed?
Fortunately, it has been shown that, at least for some cases with certain symmetry, the \ae ther vector of Einstein-\ae ther gravity can be constructed by introducing a khronon scalar $\phi$ which plays the role of time\cite{kronon1}. 
To be specific, the timelike scalar field $\phi$ determines the {\ae}ther vector\cite{BPS1,BPS2,aether1,aether2,aether3,aether4}
 \bqn
 \label{action3}
u_\mu=\frac{\partial_\mu\phi}{\sqrt{-g^{\alpha\beta}\partial_\alpha\phi\partial_\beta\phi}} .
 \eqn
In other words, to a certain degree, $\phi$ can be used to redefine physical time. 
It is further found out that Einstein-\ae ther black holes have a universal horizon which can trap any particle with arbitrarily high velocity\cite{BBM,DWW,DLWJ}.
In our recent work, we considered the khronon scalar $\phi$ and \ae ther vector $u_\mu$ as the test field in regular black hole spacetimes, and redefined time by $\phi$.
We show that the universal horizon also exists for black holes of these gravitational theories\cite{LGdW,LACW,LSWW,LSW}.

In this work, we study the static solutions of $n$ dimensional Einstein-\ae ther gravity coupling with Maxwell field as well as evaluate the universal horizon and surface gravity of those black holes.
The primary motivation of the present work to investigate the properties of higher dimensional Einstein-\ae ther gravity is due to its potential application related to gauge/gravity duality.
The latter provides a powerful toolkit for exploring a strongly coupled system via the classical gravitational theory residing in the bulk. 
Therefore, the black hole thermodynamics of the theory\cite{ho2018plb} may be associated with a lower dimensional system with strong interaction, and in particular, with broken LI.
The rest of the paper is organized as follows.
In section II, we introduce the action of EInstein-\ae ther-Maxwell theory and discuss the spherical, planar and hyperbolic solutions in asymptotically flat and (anti-)de Sitter spacetimes.
In section III, we investigate the relations between the universal horizon, killing horizon, surface horizon and electrical charge in asymptotically flat spacetime.
The conclusions are given in Section IV.
The numerical results in asymptotically (anti-)de Sitter spherical, as well as anti-de Sitter planar, spacetimes are relegated to the Appendix.

\section {Charged static solutions of Einstein-{\ae}ther theory}

According to the Einstein-\ae ther theory, the properties of the gravitational field are determined by the spacetime metric $g_{\mu\nu}$ and \ae ther vector $u^\alpha$.
In $n$-dimensional spacetime, the general action of Einstein-{\ae}ther-Maxwell theory reads\cite{BBM,DWW,DLWJ,EinsteinAether1,EinsteinAether2,EinsteinAether3,EinsteinAether4,EinsteinAether5,EinsteinAether6,EinsteinAether7,EinsteinAether8,EinsteinAether9},
 \bqn
 \label{action1}
S&=&\int d^nx \frac{\sqrt{-g}}{16\pi G_{\ae}} (R-2\Lambda+{\cal
L}_{\ae}-\alpha F_{\mu\nu}F^{\mu\nu}),
 \eqn
and
 \bqn
 \label{action2}
{\cal L}_{\ae}&=&c_1\left(\nabla_\alpha
u_\beta\right)\left(\nabla^\alpha
u^\beta\right)+c_2\left(\nabla^\beta
u_\beta\right)^2\nb\\
&&+c_3\left(\nabla_\beta u_\alpha\right)\left(\nabla^\alpha
u^\beta\right)-c_4u^\alpha u^\beta\left(\nabla_\alpha
u_\rho\right)\left(\nabla_\beta
u^\rho\right)\nb\\
&&+\lambda\left(u_\beta u^\beta+1\right),
 \eqn
where $c_1$, $c_2$, $c_3$ and $c_4$\footnote{There is a slight difference in notations from those in\cite{BBM,DWW,DLWJ}, the signs before the coefficients $c_i,\;(i=1,2,3,4)$ are inverted in the present work.} are the coupling constants in \ae ther Lagrangian ${\cal L}_{\ae}$, and $\lambda$ is Lagrange multiplier, so that $u_\alpha$ satisfies the condition $u_\beta u^\beta+1=0$.
In\cite{LGdW}, we have shown that the speed of the khronon scalar is given by
 \bqn
 \label{action4}
c_{\phi}^2=\frac{c_{123}}{c_{14}},
 \eqn
where $c_{14}\equiv c_1+c_4$, $c_{123}\equiv c_1+c_2+c_3$, and it implies that the most important cases are $c_{123}=0$ ($c_{\phi}^2=0$) and $c_{14}=0$ ($c_{\phi}^2\rightarrow\infty$).

According to the above action, one may obtain the gravitational field equation, \ae ther field equation and Maxwell equation, found in\cite{DLWJ,DWW}.

In order to investigate the charged static black hole solutions, we consider the ansatz of the solutions given by
 \bqn
 \label{metric}
ds^2&=&-F(r)dv^2+2drdv+r^2d\Sigma^2_{n-2},\nb\\
u^\alpha&=&\delta^\alpha_vu^v(r)+\delta^\alpha_rV(r)\nb\\
\zeta^\alpha&=&\delta^\alpha_v\nb\\
A_\alpha&=&\delta_\alpha^vA_0(r)
 \eqn
where $\zeta^\alpha$ is the killing vector, and we can use the condition of $u^\alpha$ to obtain $u^v=(V+\sqrt{G})/F$ with $G=F+V^2$.
The definition of $d\Sigma^2_{n-2}$ is given by
 \bqn
 \label{metric1}
d\Sigma^2_{n-2}=\left\{
  \begin{array} {ll}
  d\theta^2+\sin^2\theta d\Omega^2_{n-3}         & k=1 \nb\\
  dx_idx^i                                       & k=0 \nb\\
  d\theta^2+\sinh^2\theta d\Omega^2_{n-3}        & k=-1
\end{array}\right.
 \eqn
$k=1,0,-1$ correspondents sphere, planar and hyperboloic spacetimes respectively, and $d\Omega^2_{n-3}$ is the spherical metric.

In\cite{DLWJ}, the exact solutions and their thermodynamics for $n=3$ case was studied, so in what follows we will focus on $n\ge 4$ case.
First of all, it is easy to find the solution of the Maxwell equations:
 \bqn
 \label{m1}
A_0(r)=A_c-\frac{Qr^{3-n}}{n-3},
 \eqn
where $A_c$ and $Q$ are constant of integration in Maxwell equations.
However, it is difficult to obtain the general solutions of the gravitational equation and \ae ther equation. 
So we will only discuss the cases with $c_{14}=0$ and $c_{123}=0$ respectively.

When $c_{14}=0$ ($c_{\phi}^2\rightarrow\infty$), we find the solutions are given by
 \bqn
 \label{m2}
F(r)=F_a(r)&=&k-\frac{2M_a}{r^{n-3}}+\frac{\bar{Q}^2}{r^{2n-6}}+\frac{4c_{13}B^2}{r^{2n-4}}\nb\\
&&-\frac{2\Lambda-\rho_aB^2r_z^{2-2n}}{(n-2)(n-1)} r^2\nb\\
 V(r)=V_a(r)&=&2Br\left(r_z^{1-n}-r^{1-n}\right)
 \eqn
where
 \bqn
 \label{m3}
\bar{Q}^2&=&\frac{2\alpha Q^2}{(n-3)(n-2)}\nb\\
c_{13}&=&c_1+c_3\nb\\
\rho_a&=&4(n-1)[(n-2)c_{13}-(n-1)c_{123}]
 \eqn
and $r_z$, $M_a$ and $B$ are constants.
It is worth mentioning that $V(r_z)=0$, so $u^\alpha\propto\zeta^\alpha$ at $r=r_z$.
It is interesting to point out that when $n=4,k=1$, the above results reduce to Eq.(5.1) in\cite{DWW}.
We find that the spacetime is asymptotically de Sitter as $\Lambda>\rho_aB^2r_z^{2-2n}/2$, asymptotically anti-de Sitter as $\Lambda<\rho_aB^2r_z^{2-2n}/2$, and asymptotically flat as $\Lambda=\rho_aB^2r_z^{2-2n}/2$.

On the other hand, when $c_{123}=0$ ($c_{\phi}^2=0$), it is convenient to calculate the solutions of $F(r)$ and $G(r)$, and $V(r)$ is determined by $G=F+V^2$.
Therefore we obtain
 \bqn
 \label{m4}
F(r)=F_b(r)&=&k+\beta-\frac{2M_b}{r^{n-3}}+\frac{\tilde{Q}^2+\rho_b}{r^{2n-6}}\nb\\
&&-\frac{2\Lambda r^2}{(n-2)(n-1)(1+c_{13})} \nb\\
G(r)=G_b(r)&=&\frac{C_G}{r^{2n}}\left(r_{UH}^3r^n-r_{UH}^nr^3\right)^2
 \eqn
where
 \bqn
 \label{m5}
\tilde{Q}^2&=&\frac{2\alpha Q^2}{(n-3)(n-2)(c_{13}+1)}\nb\\
\rho_b&=&C_G\frac{(n-2)c_{13}-(n-3)c_{14}}{(n-2)(1+c_{13})r_{UH}^{-2n}}\nb\\
C_G&=&\frac{\beta+(\beta+k)c_{13}}{c_{13}r_{UH}^6}
 \eqn
and $r_{UH}$, $M_b$ and $\beta$ are constants.
$\Lambda>0$, $<0$ and $=0$ correspond to asymptotically (anti-)de Sitter and asymptotically flat spacetimes respectively.
It is noted that when $n=4,k=1,\beta=0$, the above results reduce to Eq.(5.1) in\cite{DWW}.
According to the physical content of the universal horizon, we find that $r_{UH}$ is no other than the position of the universal horizon.
What's more, refs.\cite{LGdW,LACW,LSW} have shown that the universal horizon should be between outer killing horizon and inner killing horizon in charged spacetimes, and this condition requests that universal horizon must coincide with the killing horizon at extreme black hole case, namely
 \bqn
 \label{m6}
F(r_{UH})=F'(r_{UH})=0.
 \eqn
Therefore, we have the relation
 \bqn
 \label{m7}
M_b=\left(k+\beta\right)r_{UH}^{n-3}-\frac{2\Lambda
r_{UH}^{n-1}}{(n-3)(n-1)(c_{13}+1)} .
 \eqn
In the next section, we will investigate horizons of the above spacetimes.

\section {The universal horizon and surface gravity in asymptotically flat spacetime}

As explained before, it is understood that $\phi$ is a timelike scalar, and therefore it can be used to redefine the physical time.
The universal horizon traps any particle with arbitrarily high velocity\cite{BBM,DWW,DLWJ,LGdW,LACW,LSWW,LSW}, in the sense that the geodesics cannot stretch outside of universal horizon.
It implies that black hole still exists in modified gravity with violation of Lorentz symmetry, and the Cosmic Censorship Hypothesis is still valid.
In this section, we will study the universal horizon and compare it with the killing horizon.

According to the definition of universal horizon $r_{UH}$, which satisfies
 \bqn
 \label{h1}
\left.u_\alpha\zeta^\alpha\right|_{r=r_{UH}}=0
 \eqn
The ansatz of static spacetime gives
 \bqn
 \label{h2}
G'(r_{UH})=G(r_{UH})=0.
 \eqn
and the surface gravity of universal horizon is\cite{LGdW,LACW}
 \bqn
 \label{h3}
\kappa_{UH}\equiv\left.\frac{u^a}{2}\nabla_a\left(u_\lambda\zeta^\lambda\right)\right|_{r=r_{UH}}=\left.\frac{V(r)}{2\sqrt{2}}\sqrt{G''(r)}\right|_{r=r_{UH}}.
 \eqn

On the other hand, the killing horizon $r_{KH}$ satisfies
$\left.\zeta_\alpha\zeta^\alpha\right|_{r=r_{KH}}=0$, namely
 \bqn
 \label{h4}
F(r_{KH})=0.
 \eqn
and the surface gravity at killing horizon is
 \bqn
 \label{h5}
\kappa_{KH}\equiv\left.\frac{F'(r)}{2}\right|_{r=r_{KH}}.
 \eqn
For our later convenience, we define
\bqn
r_0=(2M_a)^{1/(n-3)} .
\eqn
 
For the $c_{14}=0$ case, we can reduce 5 constants of the model ($r_0$, $r_{UH}$, $r_{KH}$, $Q$, and $\Lambda$) to 4 independent parameters ($r_U\equiv r_{UH}/r_0$, $r_k\equiv r_{KH}/r_0$, $Q^{1/(n-3)}/r_0$, and $\Lambda r_0^2$), and therefore $r_0\kappa_{KH}$ and $r_0\kappa_{UH}$ are the functions of the above new parameters.

Similarly, for the $c_{123}=0$ case, we also find that the 4 constants ($r_{UH}$, $r_{KH}$, $Q$, and $\Lambda$) can be reduced to 3 independent parameters ($r_{KH}/r_{UH}$, $Q^{1/(n-3)}/r_{UH}$, and $\Lambda r_{UH}^2$), and subsequently $r_{UH}\kappa_{KH}$ and $r_{UH}\kappa_{UH}$ are the functions of above new parameters.

In what follows, we will study asymptotically flat spherical spacetime with $k=1$ and $\Lambda=0$.
The cases for asymptotically (Anti-)de Sitter spherical spacetime and for asymptotically Anti-de Sitter planar spacetime are given in the Appendix.
We plan to study the hyperbolic ($k=-1$) case in the future.

For a black hole, the most important surface is the horizon, but the definition of the horizon can be a challenge in the gravitational theory without LI.
As discussed above, recent works\cite{universalhorizon1,universalhorizon2,universalhorizon3,universalhorizon4,universalhorizon5,universalhorizon6,universalhorizon7,universalhorizon8,universalhorizon9,universalhorizon10,universalhorizon11} show the universal horizon can be used to redefine black hole even if the gravity allows the superluminal particles.
Therefore, we will apply the above conditions to investigate the universal horizon, Killing horizon and surface gravity.

In asymptotically flat spacetime, the observer is at the infinity $r\rightarrow\infty$.
The physics in our world requires $u^r$ to satisfy the condition
$\left.u^\alpha\right|_{r\rightarrow\infty}\propto\left.\zeta^\alpha\right|_{r\rightarrow\infty}$,
in static spacetime it means
 \bqn
 \label{h6}
\left.V(r)\right|_{r\rightarrow\infty}=0
 \eqn
so we choose $r_z\rightarrow\infty$ for the $c_{14}=0$ case and $\beta=0$ for the $c_{123}=0$ case respectively.

\begin{figure*}[h]
\includegraphics[width=6cm]{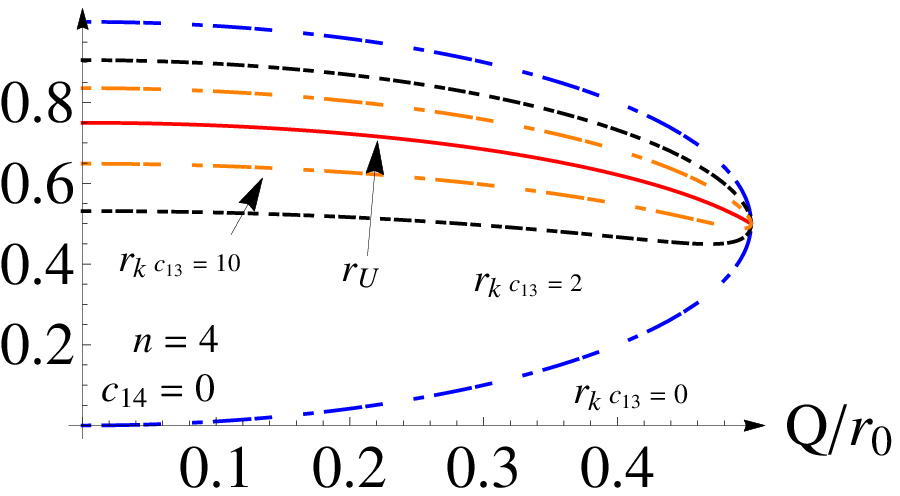}\includegraphics[width=6cm]{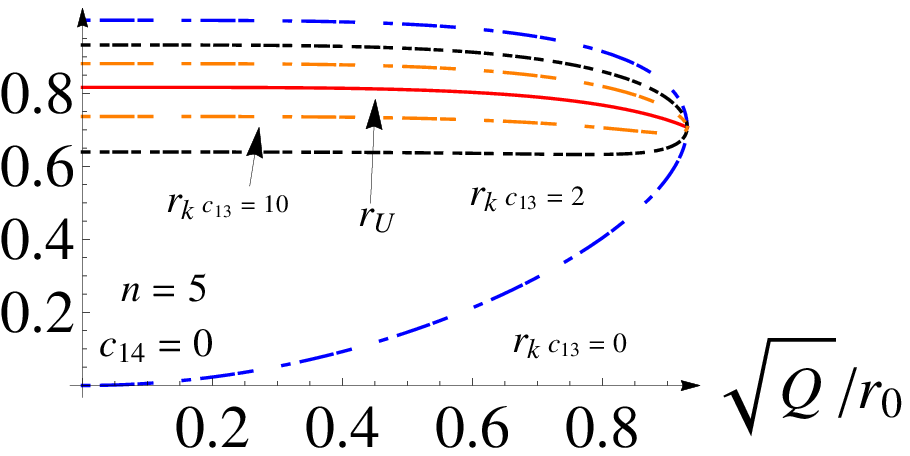}
\caption{The relation between $r_U\equiv r_{UH}/r_0$, $r_k\equiv r_{KH}/r_0$ and $Q^{1/(n-3)}/r_0$ in $4$ and $5$ dimensional asymptotically flat spherical spacetimes with $c_{14}=0$, where we have chosen $\alpha=1$.
The blue curves are for the case of $c_{13}=0$, black for $c_{13}=2$ and, orange for $c_{13}=10$.}
\label{figh0A}
\end{figure*}

\begin{figure*}[h]
\includegraphics[width=6cm]{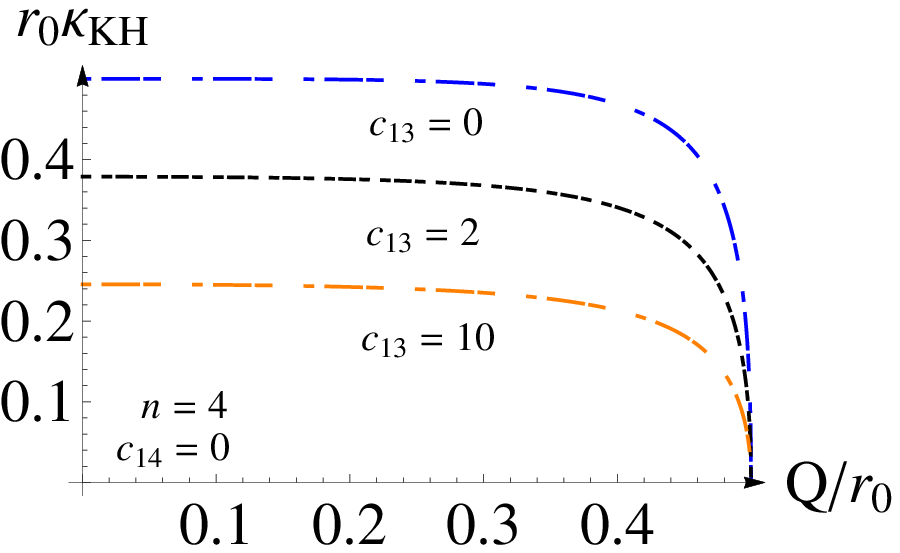}\includegraphics[width=6cm]{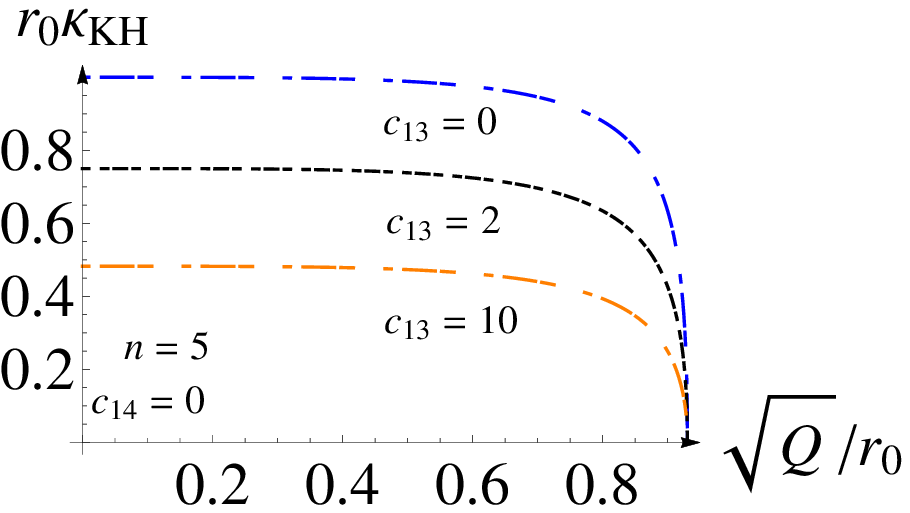}
\includegraphics[width=6cm]{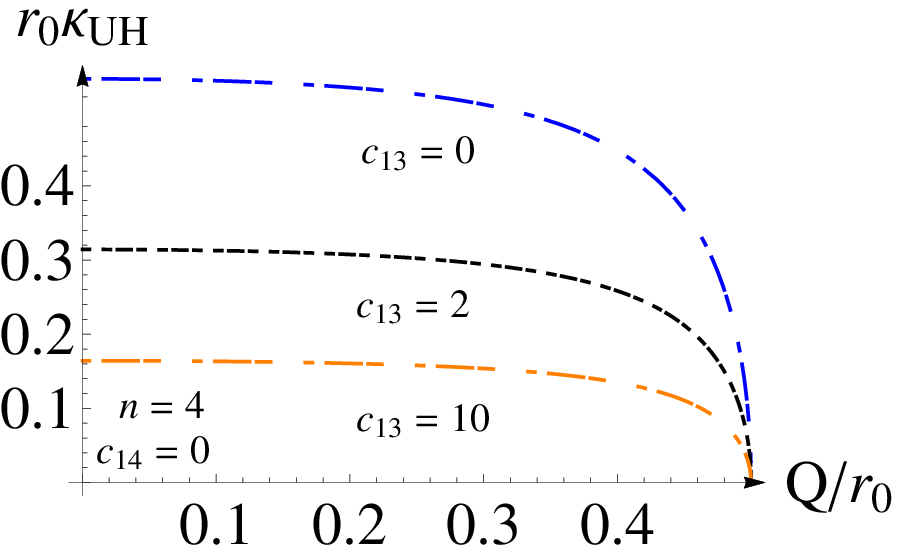}\includegraphics[width=6cm]{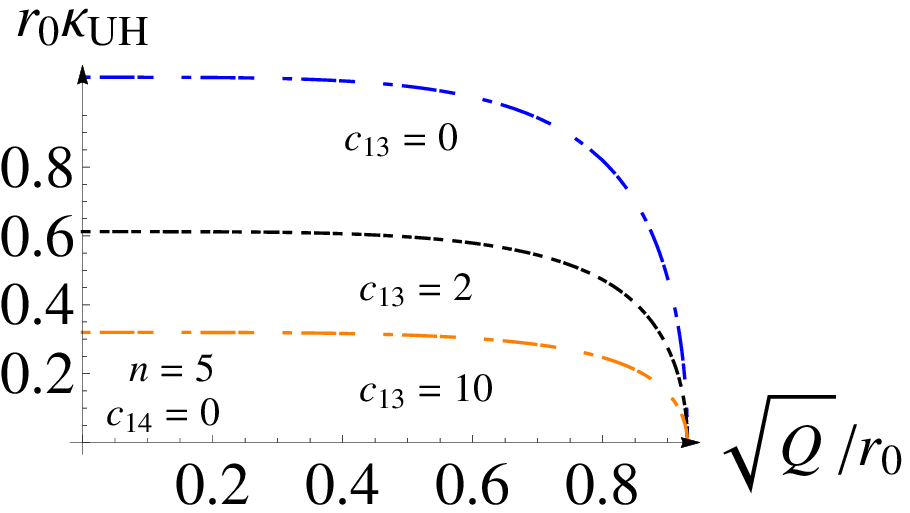}
\caption{The relation between $r_0\kappa_{KH}$, $r_0\kappa_{UH}$ and
$Q^{1/(n-3)}/r_0$ in $4$ and $5$ dimensional asymptotically flat
spherical spacetimes for various $c_{13}>0$ with $c_{14}=0$, $\alpha=1$.} \label{figk0A}
\end{figure*}

\begin{figure*}[h]
\includegraphics[width=6cm]{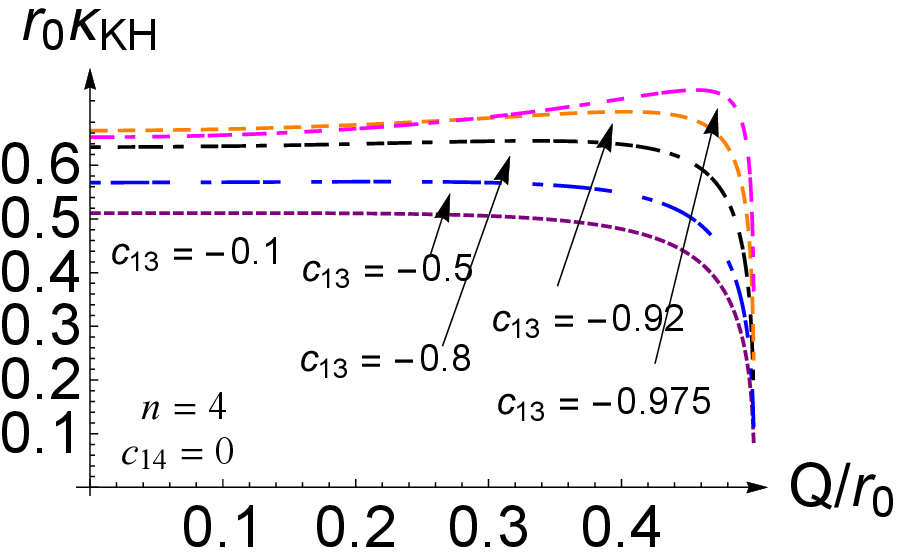}\includegraphics[width=6cm]{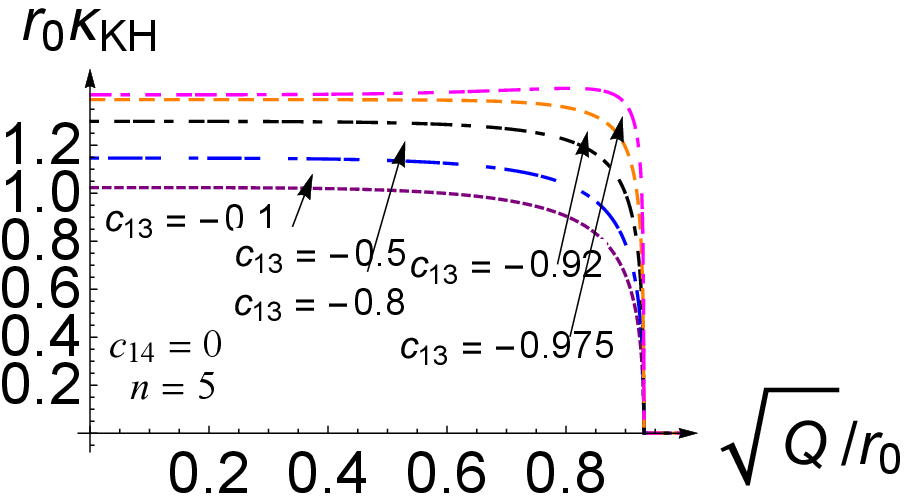}
\includegraphics[width=6cm]{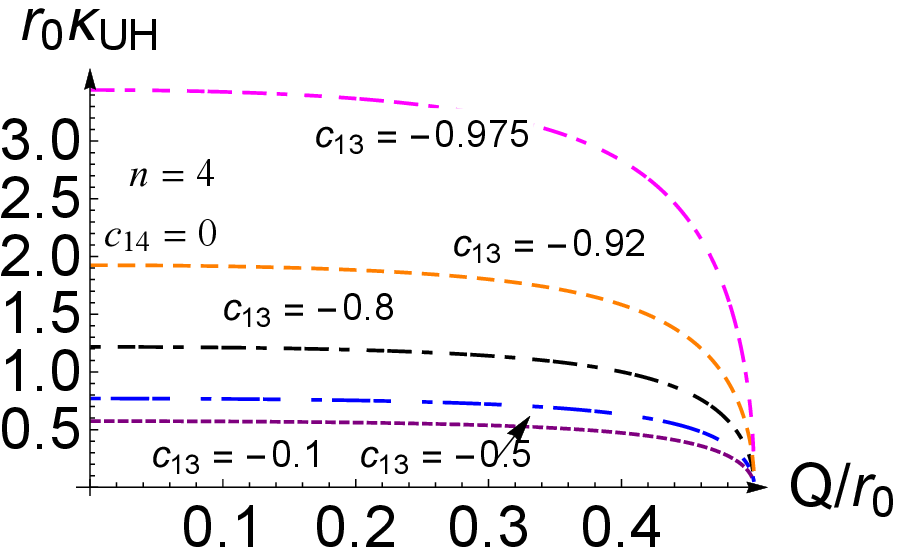}\includegraphics[width=6cm]{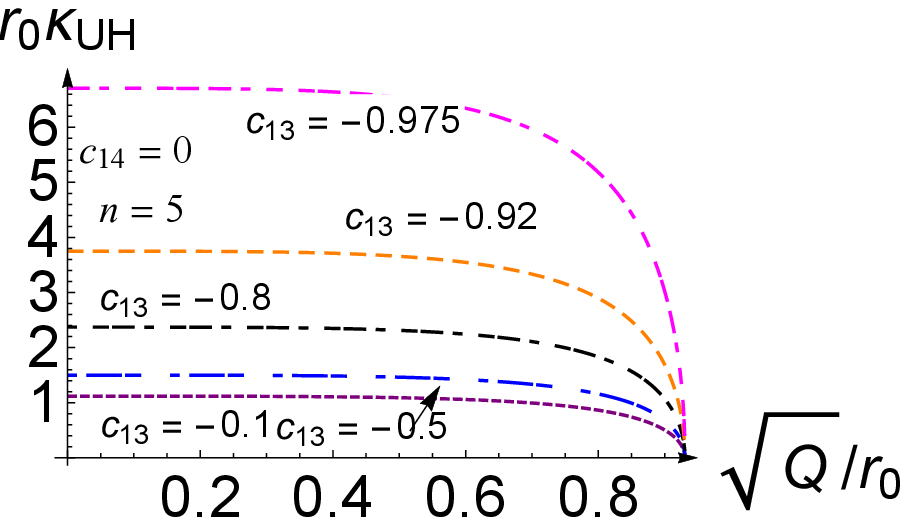}
\caption{The relation between $r_0\kappa_{KH}$, $r_0\kappa_{UH}$ and
$Q^{1/(n-3)}/r_0$ in $4$ and $5$ dimensional asymptotically flat
spherical spacetimes for various $-1<c_{13}<0$ with $c_{14}=0$, $\alpha=1$.} \label{xinA}
\end{figure*}

\begin{figure*}[h]
\includegraphics[width=6cm]{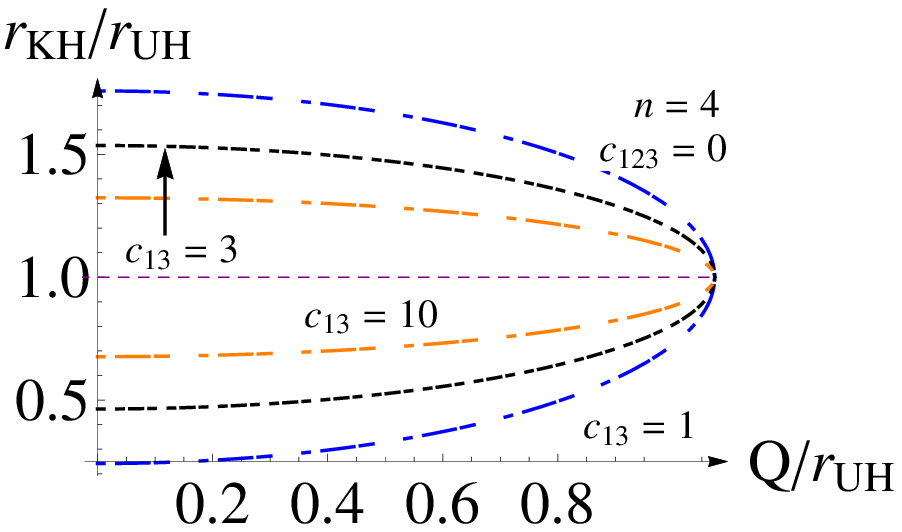}\includegraphics[width=6cm]{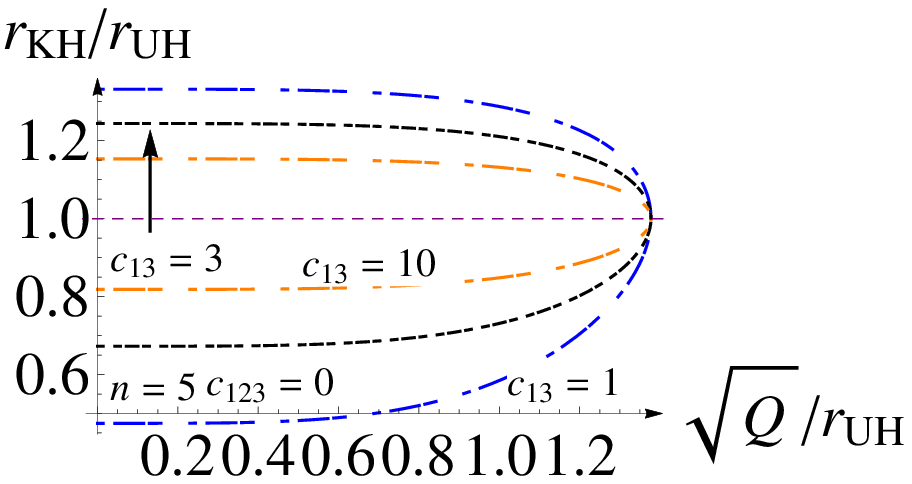}
\caption{The relation between $r_{KH}/r_{UH}$ and $Q^{1/(n-3)}/r_0$ in $4$ and $5$ dimensional asymptotically flat spherical spacetimes with $c_{123}=0$, where we have chosen $\alpha=1$ and $c_{14}=0.3$. The blue curves are for the case of $c_{13}=1$, black for $c_{13}=3$ and, orange for $c_{13}=10$.}
\label{figh0B}
\end{figure*}

\begin{figure*}[h]
\includegraphics[width=6cm]{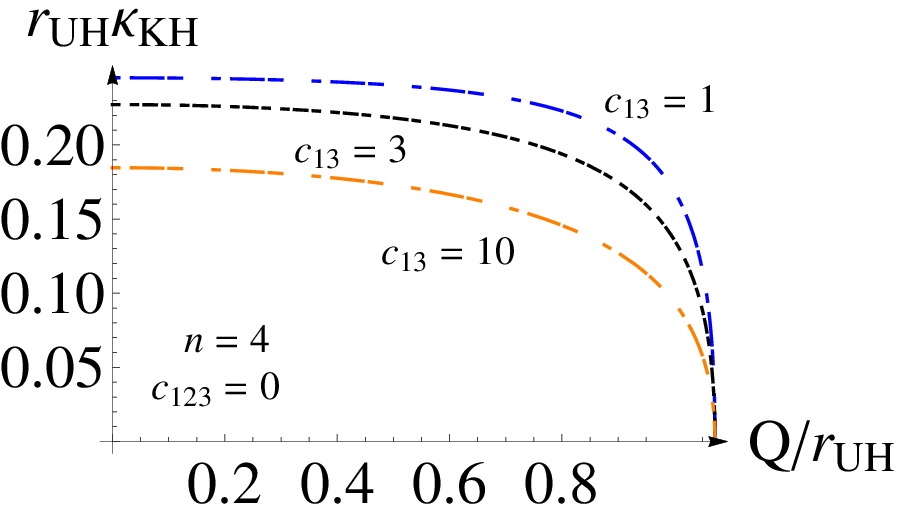}\includegraphics[width=6cm]{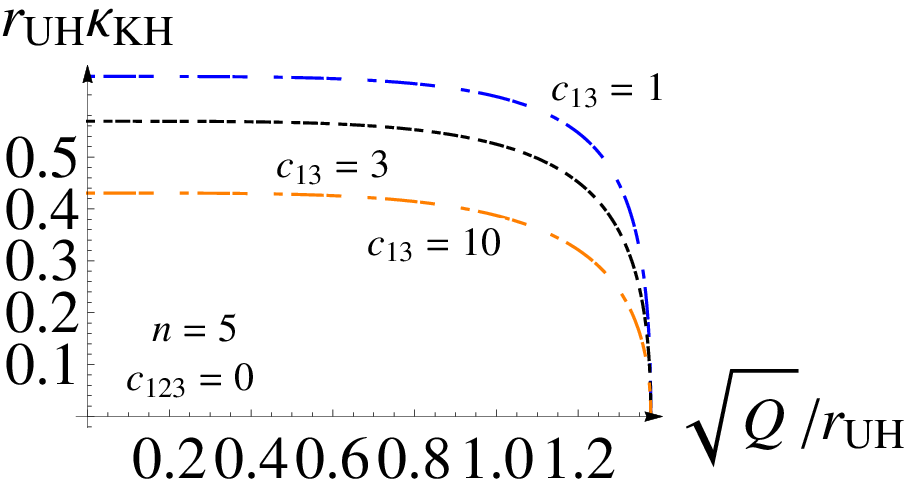}
\includegraphics[width=6cm]{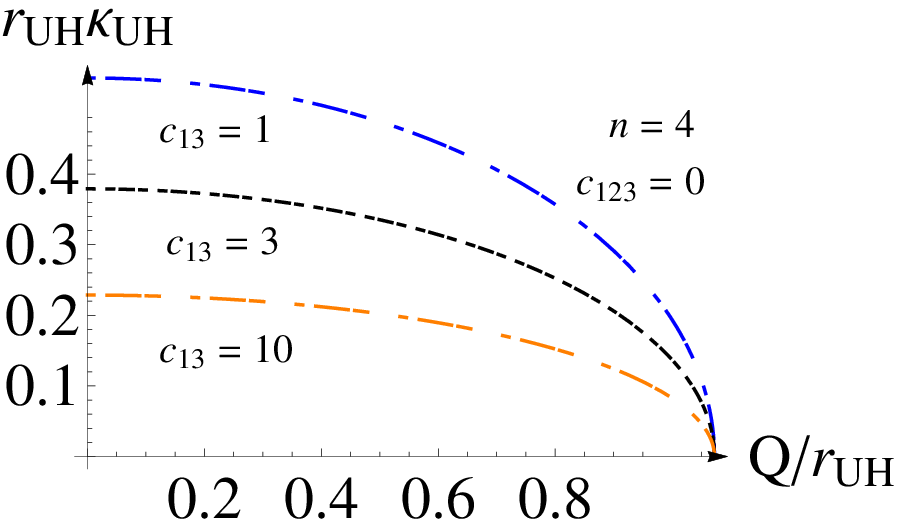}\includegraphics[width=6cm]{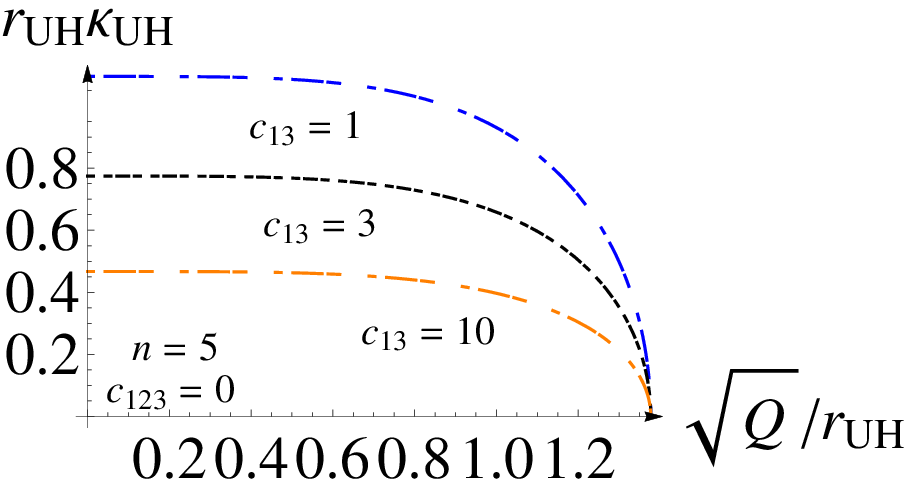}
\caption{The relation between $r_{UH}\kappa_{KH}$, $r_{UH}\kappa_{UH}$ and $Q^{1/(n-3)}/r_0$ in $4$ and $5$ dimensional asymptotically flat spherical spacetimes with $c_{123}=0$ with $c_{13}>0$, $\alpha=1$ and $c_{14}=0.3$.} \label{figk0B}
\end{figure*}

We show the relations between the horizon and the surface gravity in Figs.\ref{figh0A}-\ref{figk0B}.

In Fig.\ref{figh0A}, the three curves above $r_U$ are the outer Killing horizons $r_h$ and, the others three represent the inner Killing horizons $r_i$.
Therefore, $r_h$, $r_i$ and the universal horizon satisfy the relation $r_i\le r_{UH}\le r_h$, and they coincide for the case of the extreme black hole, as expected.
Two Killing horizons are found to approach the universal horizon as $c_{13}$ increases. It implies that universal horizon would also coincide with outer and inner Killing horizon as $c_{13}\rightarrow\infty$.

Fig.\ref{figk0A} and \ref{xinA} show the results of surface gravity, and we find that the surface gravity becomes smaller as $c_{13}$ increases.
The figures indicate that the surface gravity disappears as $c_{13}\rightarrow\infty$, and this is where the three horizons coincide.
The calculations are carried out at Killing and universal horizons, it is shown that $\kappa_{KH}$ and $\kappa_{UH}$ vanish for the case of extreme black holes.
For $c_{13}>0$, the product $r_0\kappa_{\mathrm KH}$ decreases with increasing $Q^{1/(n-3)}/r_0$ until it vanishes at the limit of extreme black hole.
However, for $c_{13 }<0$, especially in the region where $c_{13}\rightarrow -1$, $r_0\kappa_{\mathrm KH}$ first increases and then decrease abruptly and vanishes.
We note that this feature is found to depend on the dimension.

Similarly, for the $c_{123}=0$ case with fixed $c_{14}$, from Fig.\ref{figh0B}, we find that outer and inner horizon approach the universal horizon as $c_{13}$ increases, and it implies that the three horizons will coincide as $c_{13}\rightarrow\infty$.

\begin{figure*}[h]
\includegraphics[width=6cm]{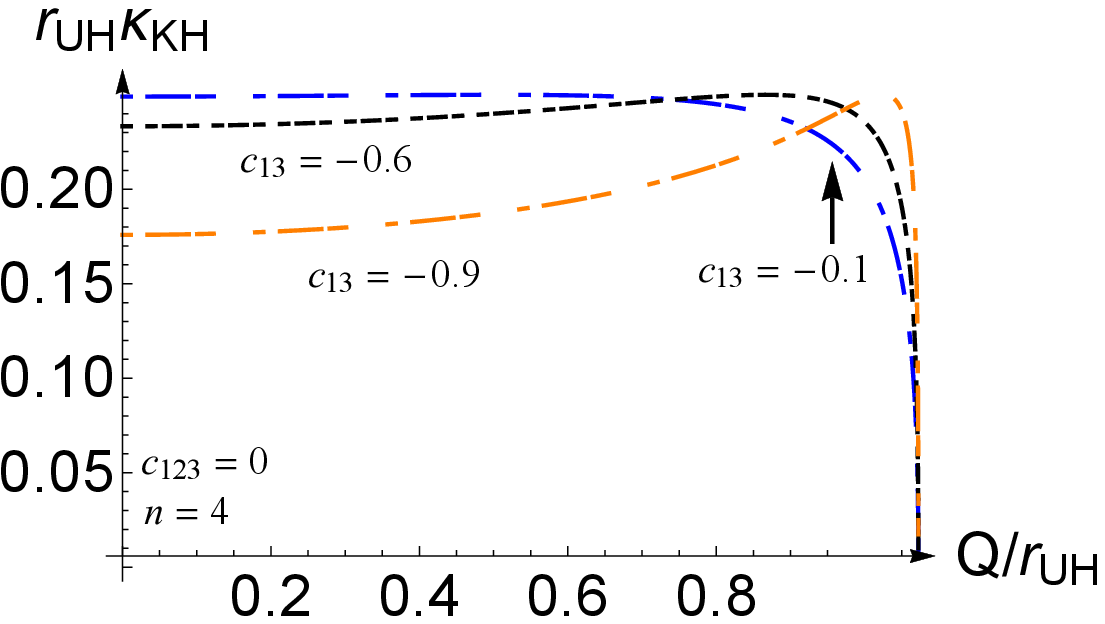}\includegraphics[width=6cm]{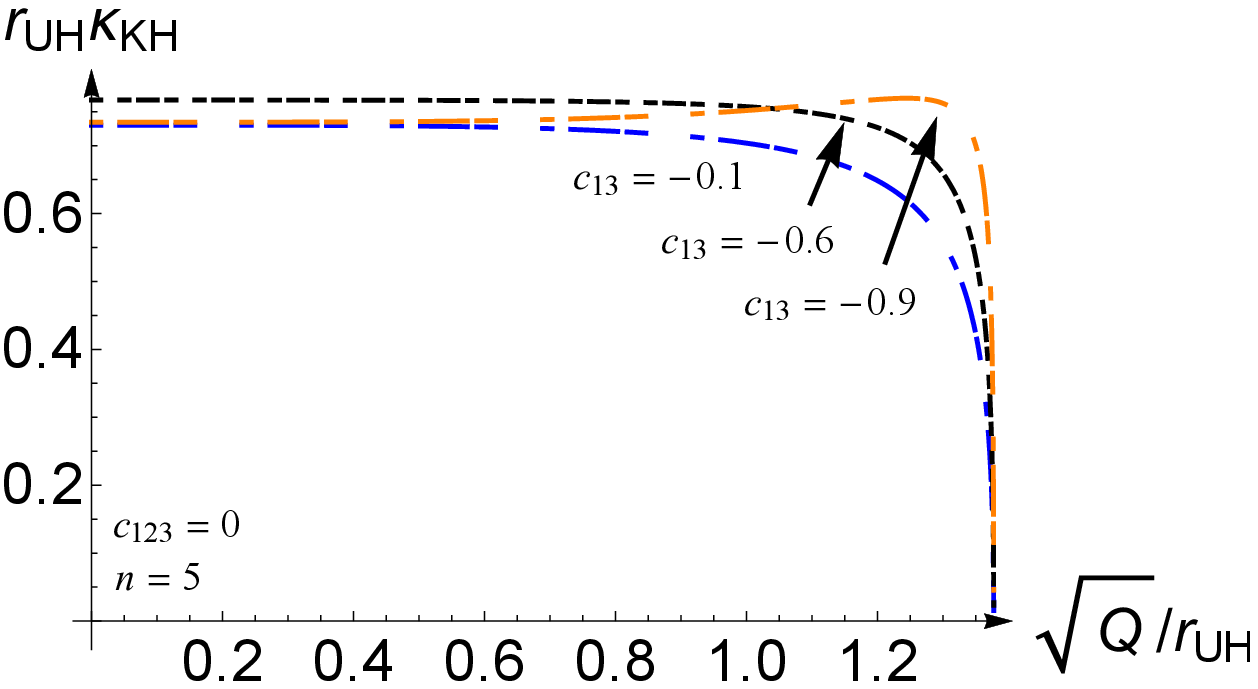}
\includegraphics[width=6cm]{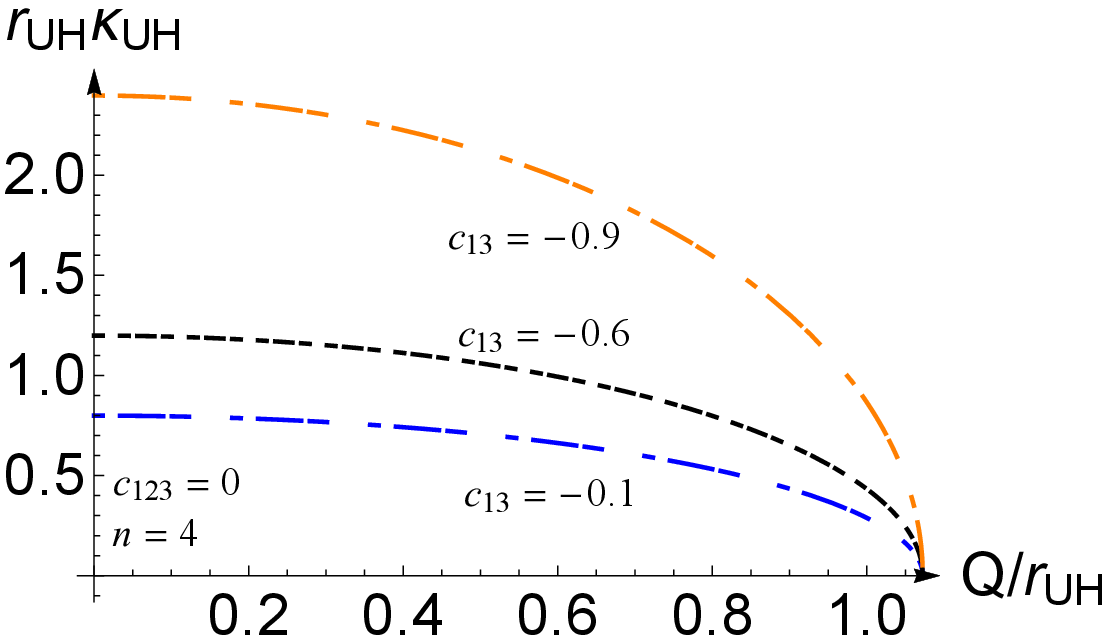}\includegraphics[width=6cm]{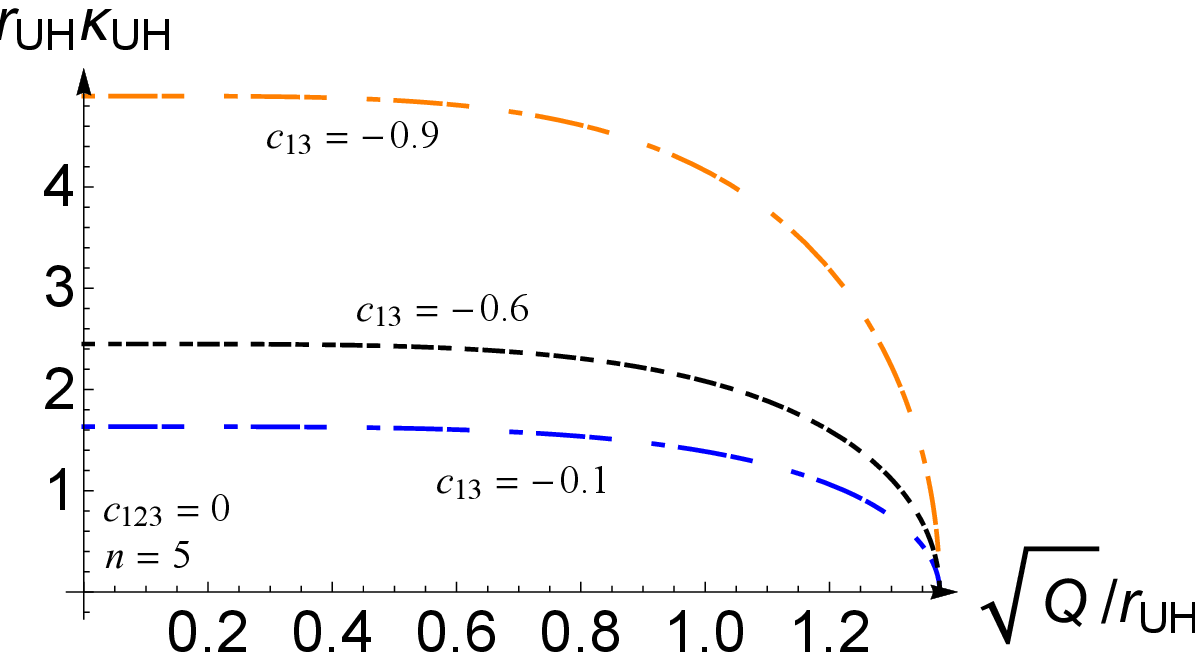}
\caption{The relation between $r_{UH}\kappa_{KH}$, $r_{UH}\kappa_{UH}$ and $Q^{1/(n-3)}/r_0$ in $4$ and $5$ dimensional asymptotically flat spherical spacetimes with $c_{123}=0$ with $-1<c_{13}<0$, $\alpha=1$ and $c_{14}=0.3$.} \label{xinB}
\end{figure*}

The Fig.\ref{figk0B} and \ref{xinB} also show similar behavior with the surface gravity for the $c_{14}=0$ case: the larger $c_{13}$ leads to smaller $\kappa_{KH}$ and $\kappa_{UH}$, and it indicates that surface gravities vanish as $c_{13}\rightarrow\infty$.
For $c_{13}>0$, the product $r_{\mathrm UH}\kappa_{\mathrm UH}$ monotonically decreases with increasing $Q^{1/(n-3)}/r_{\mathrm UH}$ until it vanishes at the limit of the extreme black hole.
However, for $c_{13 }<0$, particularly for the region where $c_{13}\rightarrow -1$, $r_{\mathrm UH}\kappa_{\mathrm UH}$ first increases slowly and then decrease rapidly.
Again, this feature is found to depend on the dimension.

\begin{figure*}[h]
\includegraphics[width=6cm]{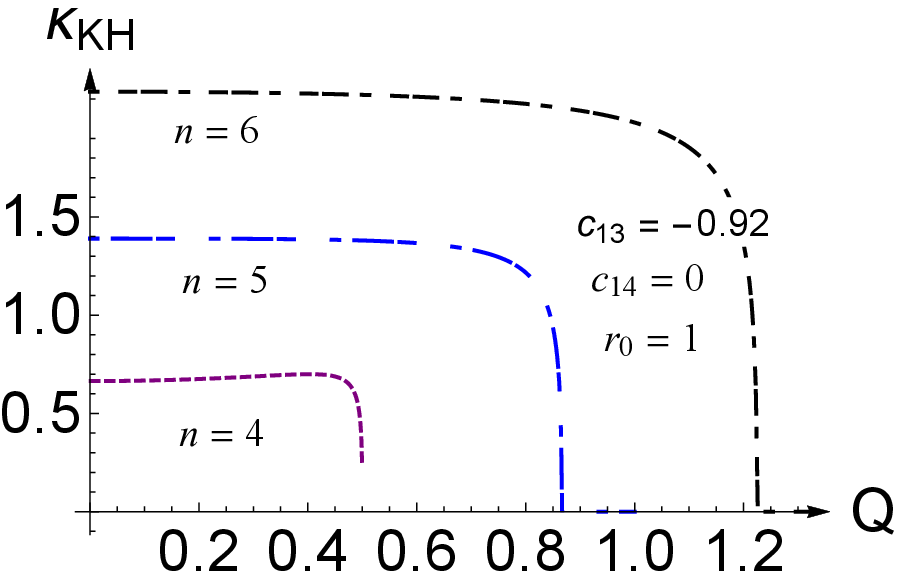}\includegraphics[width=6cm]{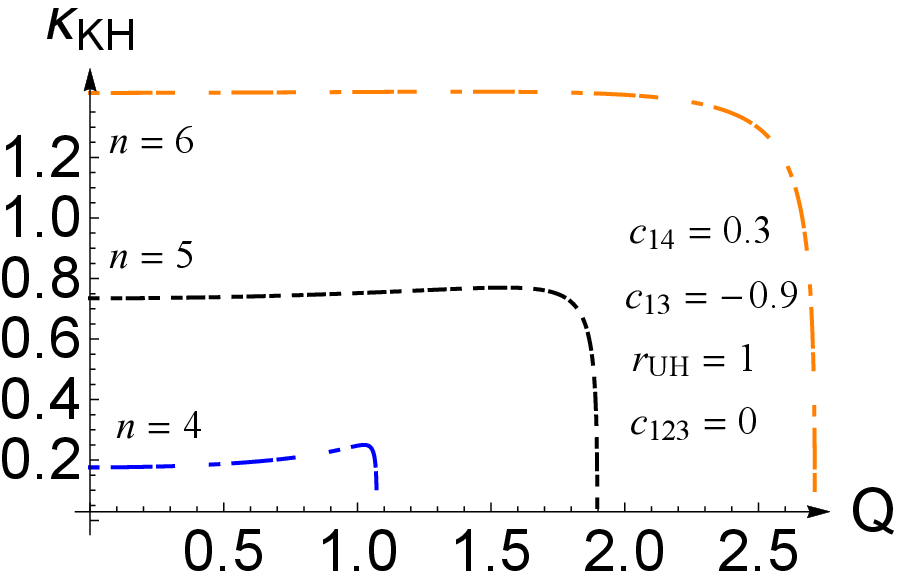}
\includegraphics[width=6cm]{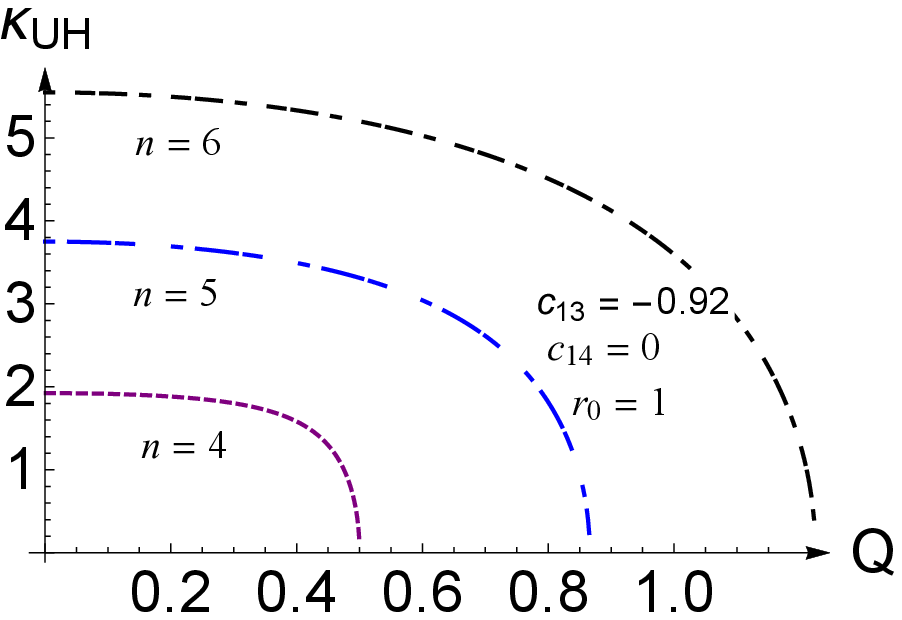}\includegraphics[width=6cm]{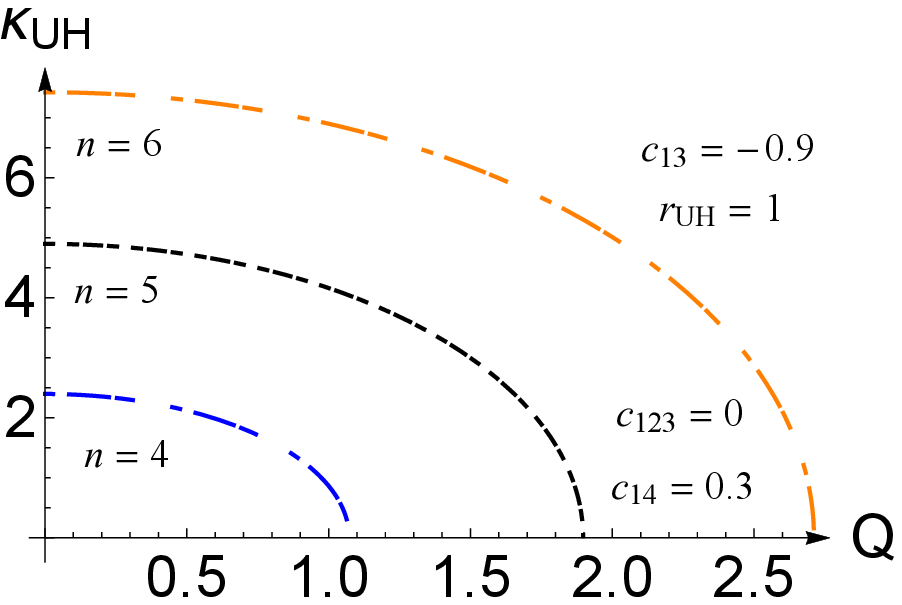}
\caption{The relation between $\kappa_{KH}$, $\kappa_{UH}$ and $Q$ for various dimensions $n=4,5$ and $6$ in asymptotically flat spherical spacetimes.} \label{NN}
\end{figure*}
In order to explicitly show the dimension dependence, for given parameters we present in Fig.\ref{NN} the results by various only the dimension $n$.
It is found that as $n$ increases, the transit to the extreme black hole at the horizon becomes slightly more gradual. 
For the Killing horizon, it becomes a monotonically decreasing function of $Q$ as $n$ reaches $6$.

The main charactersitcs of the universal and killing horizons, as well as how they transit to the case of the extreme black hole, are mostly found to be similar to those in the lower dimensional cases\cite{BBM,DWW,DLWJ}.
The difference, if any, seems to be quantitative.
On the other hand, the calculated surface gravity, for some specific choice of parameters (for instance when $c_{14}=0$ and $c_{13}\rightarrow -1$), the dimensional dependence of the results becomes more substantial.
The difference is more observable as the solution transits to that of the extreme black hole.

\section{Conclusion}

In this work, we investigated the black holes solutions of n-dimensional Einstein-\ae ther-Maxwell theory, and then analyzed the behavior of the killing and the universal horizons.
The results of the present work can be viewed as a generalization of previous studies\cite{BBM,DWW,DLWJ}.
The latter can be readily restored if one substituting $c_{14}=0$ in the solution shown in Eq.(\ref{m2}) and $c_{123}=0$ in Eq.(\ref{m4}), while taking $k=1$, $\Lambda=0$, $\beta=0$, and $r_z\rightarrow\infty$.
Our study reveals that the universal horizon continues to exist in the higher dimensional black hole spacetimes.
It is found that the main features concerning the horizon, as well as surface gravity, are mostly similar to those in $n=3$ case.
Some subtle difference is observed, for instance, the transition to extreme black hole becomes more smooth as the dimension $n$ increases.

Our calculations further confirm that, in the higher dimensional modified gravitational theory in question, the black hole solution still possesses the extreme case, which is defined by Eq.(\ref{m6}).
Moreover, the universal horizon coincides with the killing horizon at such extreme condition while the surface gravities $\kappa_{UH}$ and $\kappa_{KH}$ vanish.
Mathematically, one has $F(r_{UH})=G(r_{UH})=0$ at the extreme case, and it requires $V(r_{UH})=0$, so $\kappa_{UH}=V(r_{UH})\sqrt{G''(r_{UH})}/(2\sqrt{2})=0$.
This result implies that the black hole's temperature $T=\kappa/2\pi$ equals $0$ at such extreme case.
If one can prove that ``It is not possible to form a black hole with vanishing Temperature", (or ``vanishing surface gravity is not possible to achieve"), it signifies that the Third Law of black hole thermodynamics applies to the case of black holes with a universal horizon.
When the electrical charge $Q$ is larger than $Q_e$ of the extreme case, one has $F>0$, and $G=F+V^2>0$, so that the universal horizon vanishes, which would result in a Naked singularity.
It is prohibited by the Cosmic Censor Conjecture\cite{cosmo}.

Moreover, in higher dimensional geometry, one can still express the metric as
 \bqn
 \label{f1}
g_{ab}=-u_au_b+s_as_b+\hat{g}_{ab}
 \eqn
where $\hat{g}_{ab}\equiv(0,0,g_{ii})$, and the spacelike unit vector $s_a$ satisfties $u^as_a=0$ and $s^2=1$.
One can use this expression to discuss the Smarr Formula and the first law of thermodynamics in higher dimensional charged Einstein-{\ae}ther black hole spacetimes.
Also, few real black holes in our universe might be absolutely stationary, so it is essential to investigate rotational black hole solutions, and we plan to carry out these studies in the near further.

\section*{Acknowledgments}

We are thankful for Prof. Anzhong Wang and Prof. Chikun Ding for valuable discussions and insightful comments.
We gratefully acknowledge the financial support from
Funda\c{c}\~ao de Amparo \`a Pesquisa do Estado de S\~ao Paulo (FAPESP),
Funda\c{c}\~ao de Amparo \`a Pesquisa do Estado do Rio de Janeiro (FAPERJ),
Conselho Nacional de Desenvolvimento Cient\'{\i}fico e Tecnol\'ogico (CNPq),
and Coordena\c{c}\~ao de Aperfei\c{c}oamento de Pessoal de N\'ivel Superior (CAPES).
A part of the work was developed under the National Natural Science Foundation of China (NNSFC) under contract No.11805166, 11573022 and 11375279.

\appendix
\section{Asymptotically (Anti-)de Sitter spherical spacetime and Anti-de Sitter planar spacetime}

\subsection {Asymptotically de Sitter spherical spacetime \\$k=1$ and $\Lambda>0$}

In asymptotically de Sitter spacetimes, the position $r\rightarrow\infty$ is a singularity, and nobody can stay at this point, so the condition $\left.u^\alpha\right|_{r\rightarrow\infty}\propto\left.\zeta^\alpha\right|_{r\rightarrow\infty}$ is not necessary.
It means that we don't need the conditions $r_z\rightarrow\infty$ for the $c_{14}=0$ case and $\beta=0$ for the $c_{123}=0$ case.

In charged de Sitter spacetimes, black holes have three killing horizons: outer killing horizon $r_{p}$, inner killing horizon $r_i$ and cosmological killing horizon $r_C$, and the relation between the killing horizon and the universal horizon is given by $r_i<r_{UH}<r_p<r_C$.

\begin{figure*}[h]
\includegraphics[width=6cm]{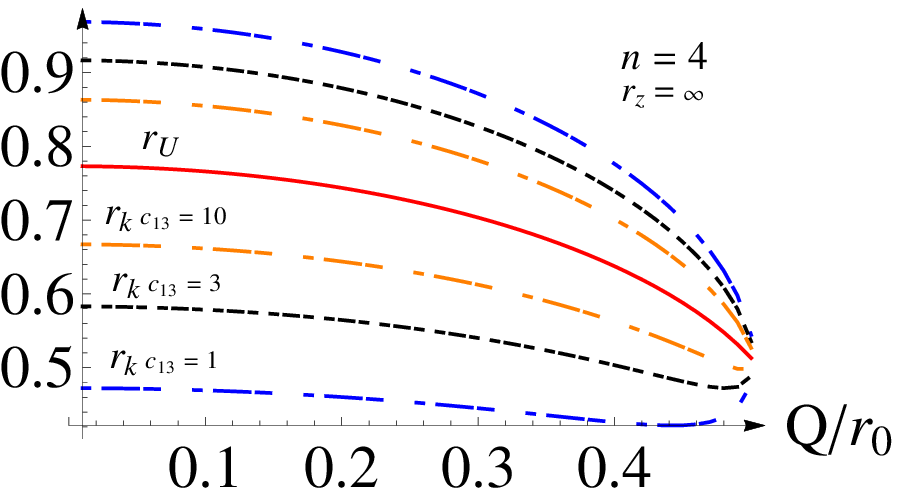}\includegraphics[width=6cm]{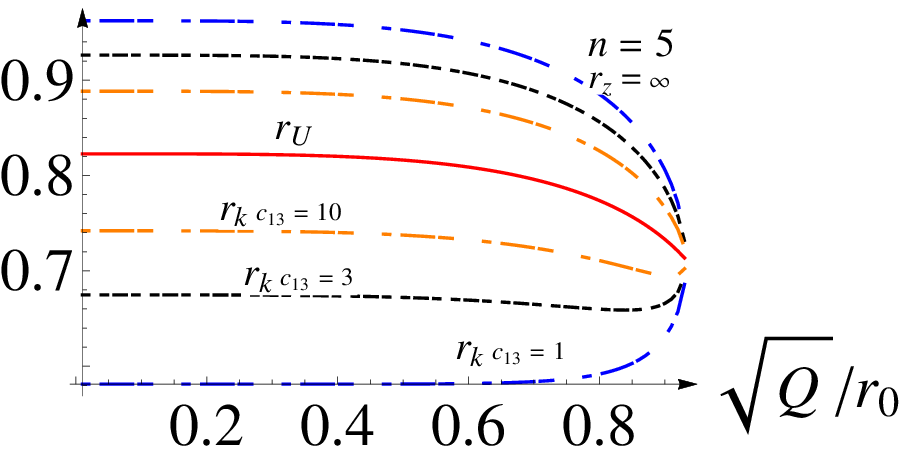}
\caption{The relation between $r_{KH}/r_{UH}$ and $Q^{1/(n-3)}/r_0$ in $4$ and $5$ dimensional asymptotically de Sitter spherical spacetimes with $c_{14}=0$, where we have chosen $\Lambda=0.1/r_0^2$, $\alpha=1$ and $c_{123}=3$.} \label{fighPA}
\end{figure*}

\begin{figure*}[h]
\includegraphics[width=6cm]{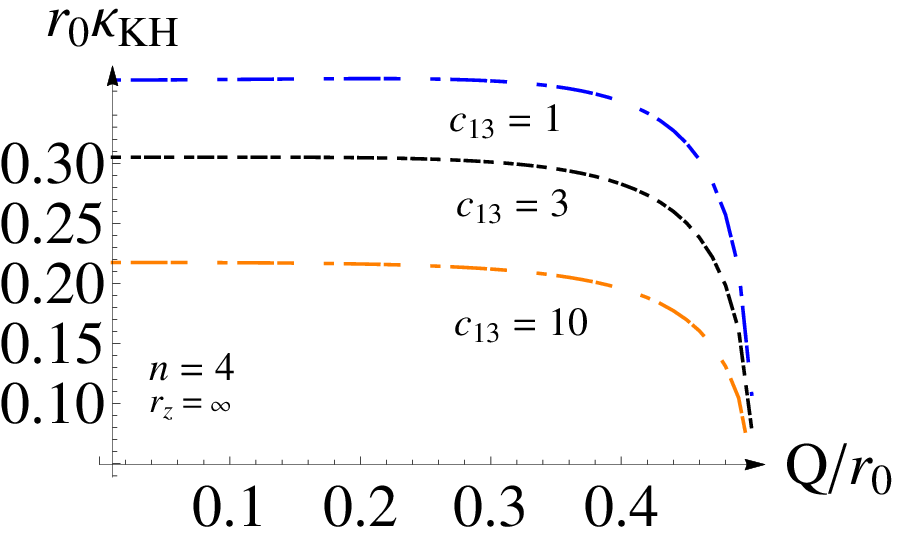}\includegraphics[width=6cm]{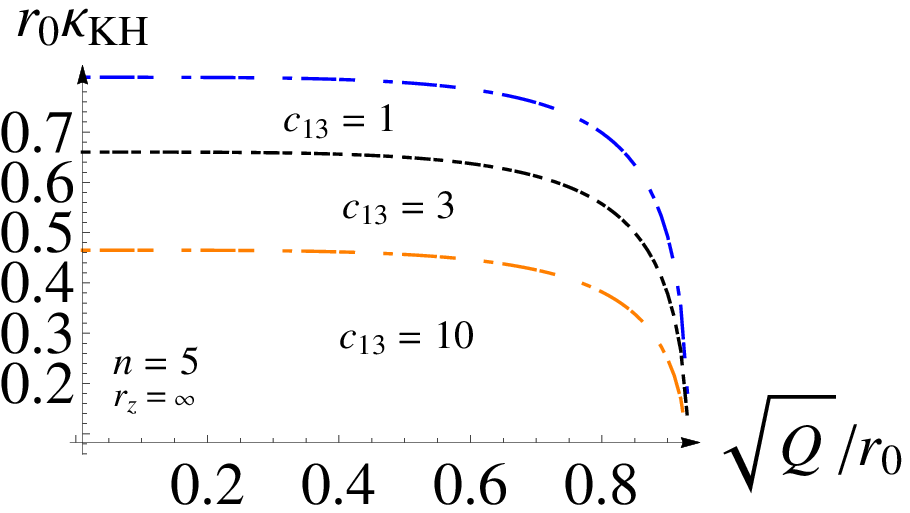}
\includegraphics[width=6cm]{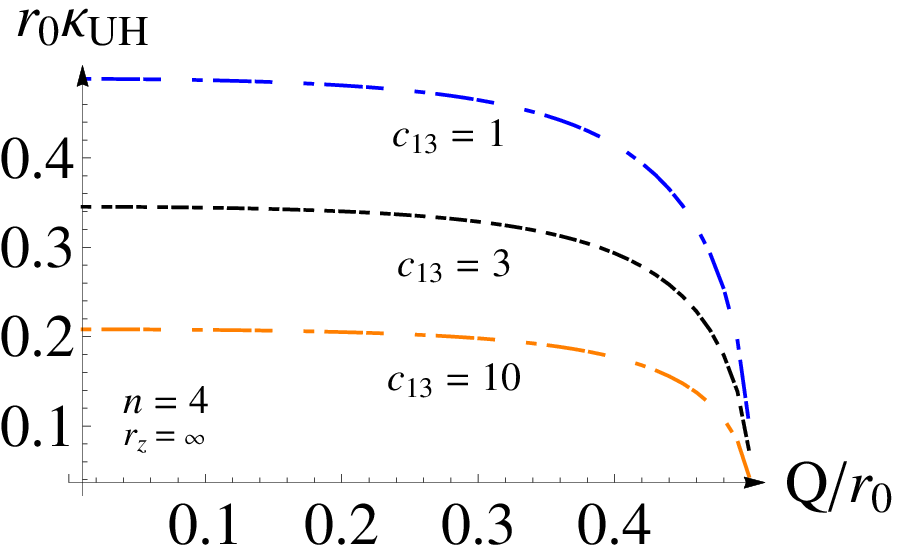}\includegraphics[width=6cm]{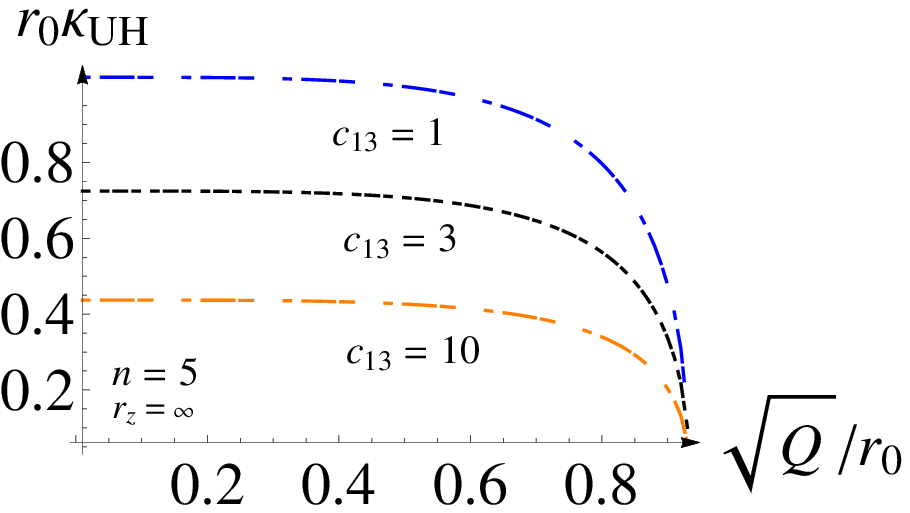}
\caption{The relation between $r_0\kappa_{KH}$, $r_0\kappa_{UH}$ and
$Q^{1/(n-3)}/r_0$ in $4$ and $5$ dimensional asymptotically de Sitter spherical spacetimes with $c_{14}=0$, where we have chosen $\Lambda=0.1/r_0^2$, $\alpha=1$ and $c_{123}=3$.} \label{figkPA}
\end{figure*}

\begin{figure*}[h]
\includegraphics[width=6cm]{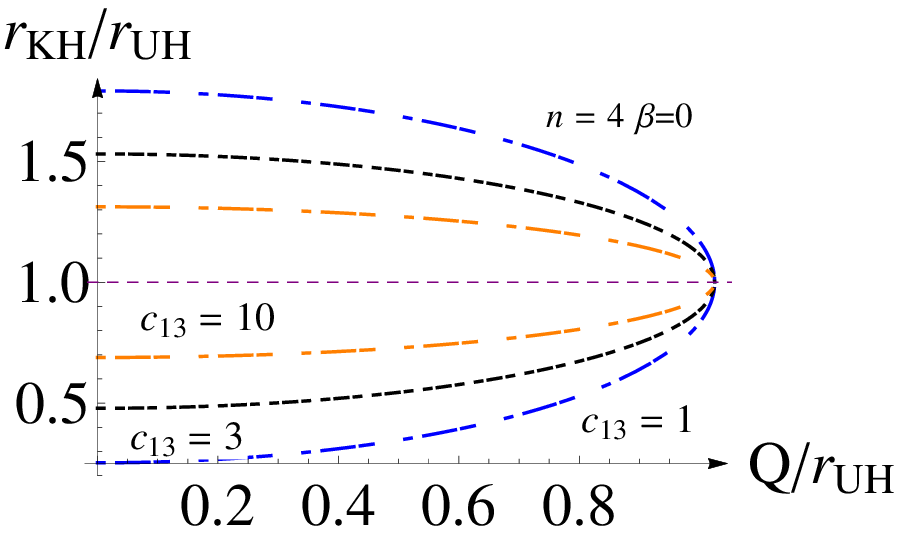}\includegraphics[width=6cm]{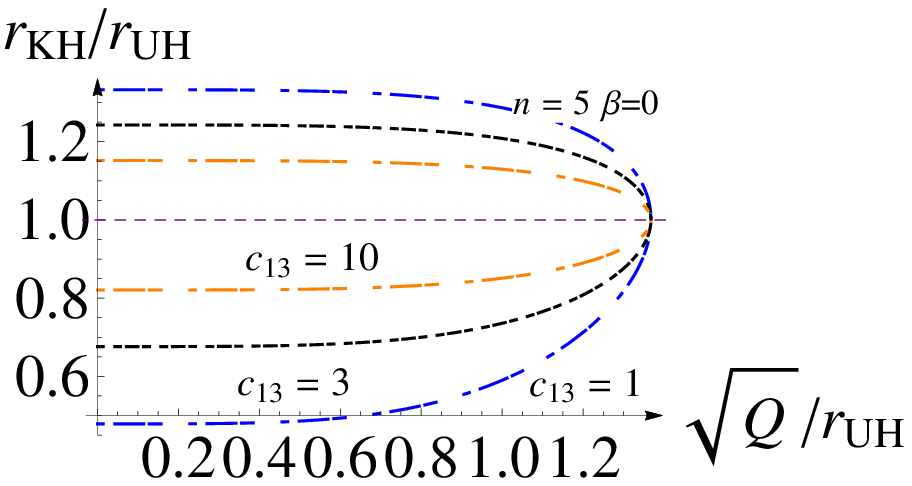}
\caption{The relation between $r_{KH}/r_{UH}$ and $Q^{1/(n-3)}/r_{UH}$ in $4$ and $5$ dimensional asymptotically de Sitter spherical spacetimes with $c_{123}=0$, where we have chosen $\Lambda=0.1/r_0^2$, $\alpha=1$ and $c_{14}=0.3$.} \label{fighPB}
\end{figure*}

\begin{figure*}[h]
\includegraphics[width=6cm]{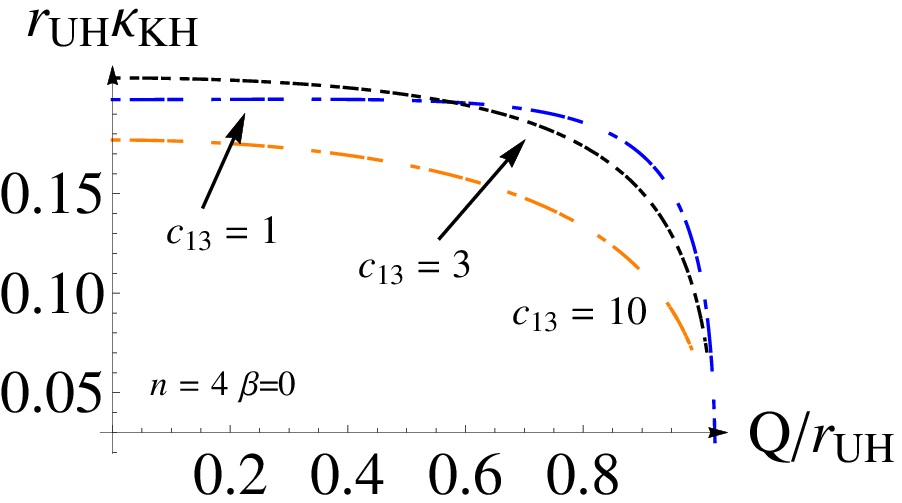}\includegraphics[width=6cm]{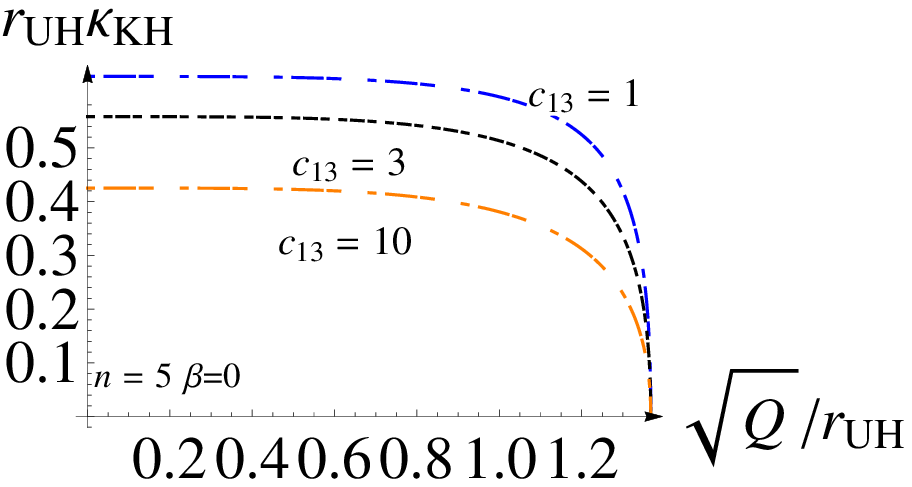}
\includegraphics[width=6cm]{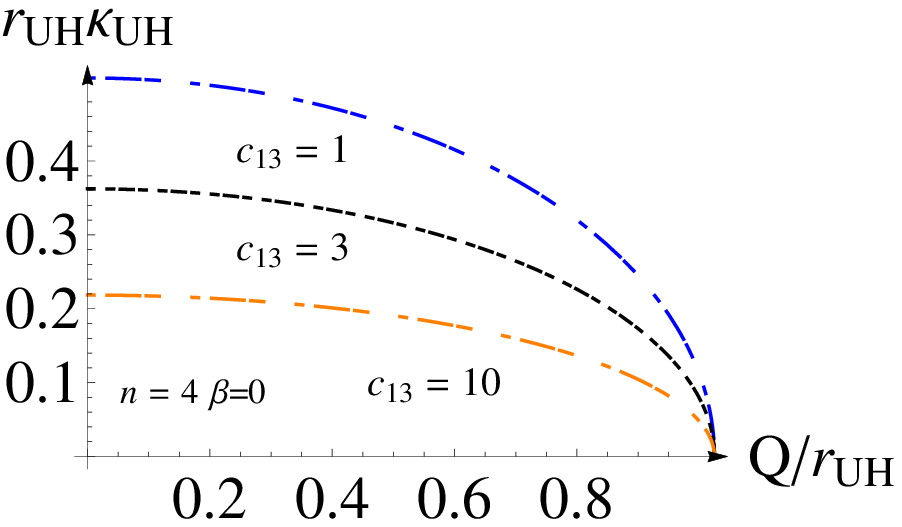}\includegraphics[width=6cm]{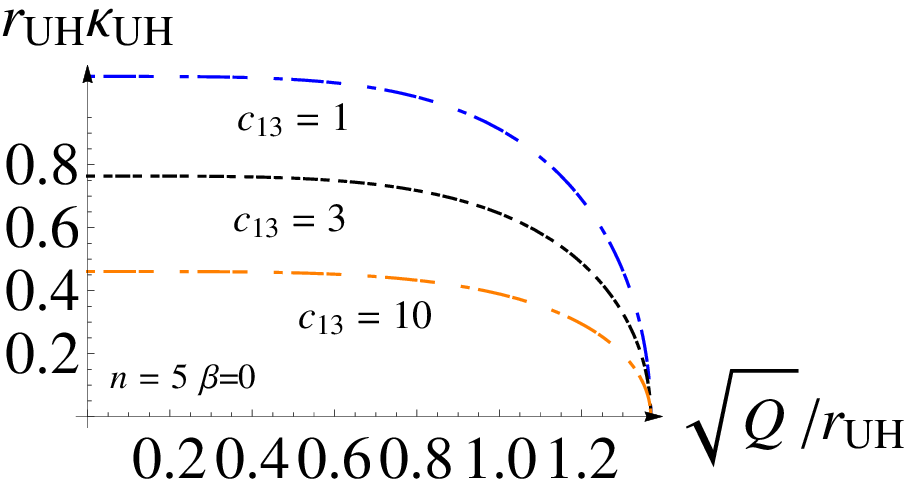}
\caption{The relation between $r_{UH}\kappa_{KH}$, $r_{UH}\kappa_{UH}$ and $Q^{1/(n-3)}/r_{UH}$ in $4$ and $5$ dimensional asymptotically de Sitter spherical spacetimes with $c_{123}=0$, where we have chosen $\Lambda=0.1/r_0^2$, $\alpha=1$ and $c_{14}=0.3$.} \label{figkPB}
\end{figure*}

We show the relation between horizon and surface gravity with $r_z\rightarrow\infty$ and $\beta=0$ in Figs.\ref{fighPA}-\ref{figkPB}.

\subsection {Asymptotically Anti-de Sitter spherical spacetime \\$k=1$ and $\Lambda<0$}

Anti-de Sitter spacetime implies negative $\Lambda$, and this spacetime plays a significant role in the research of modern theoretical physics, owing to the correspondence between Anti-de Sitter and conformal field theory (AdS/CFT correspondence).
Therefore, it is meaningful to study the Anti-de Sitter black hole in Einstein-\ae ther theory.

\begin{figure*}[h]
\includegraphics[width=6cm]{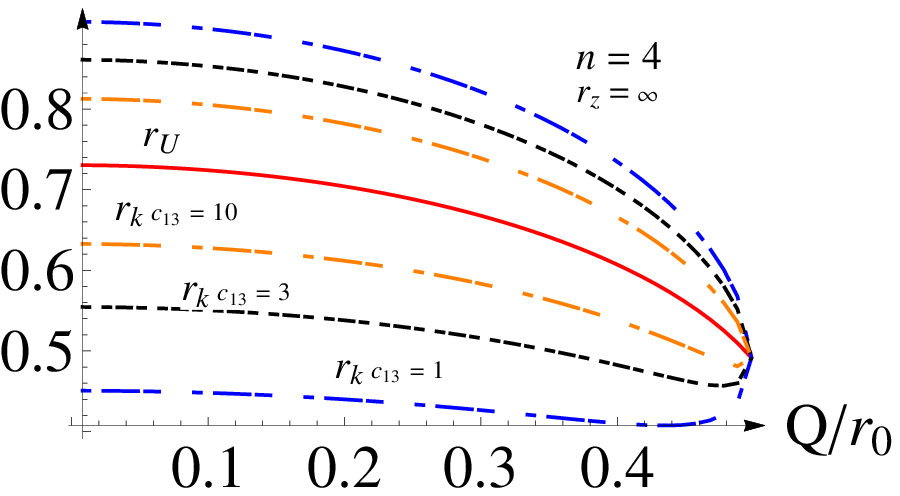}\includegraphics[width=6cm]{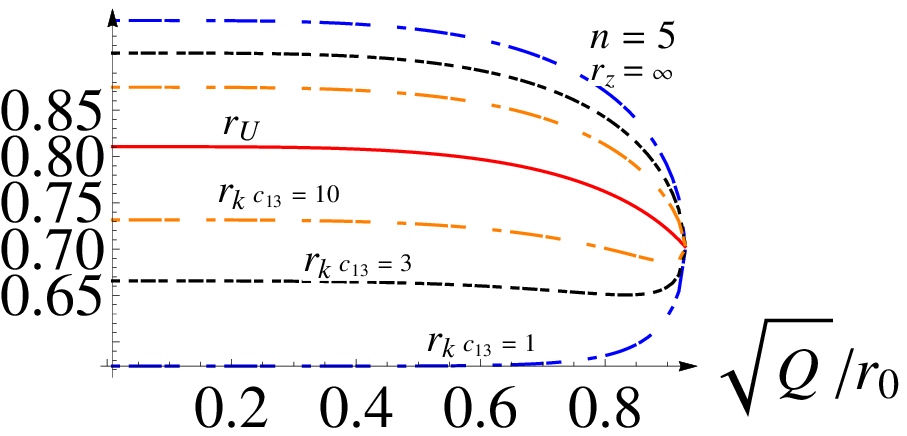}
\caption{The relation between $r_U\equiv r_{UH}/r_0$, $r_k\equiv
r_{KH}/r_0$ and $Q^{1/(n-3)}/r_0$ in $4$ and $5$ dimensional
asymptotically Anti-de Sitter spherical spacetimes with $c_{14}=0$,
where we have chosen $\Lambda=-0.1/r_0^2$, $\alpha=1$ and
$c_{123}=3$.} \label{fighNA}
\end{figure*}

\begin{figure*}[h]
\includegraphics[width=6cm]{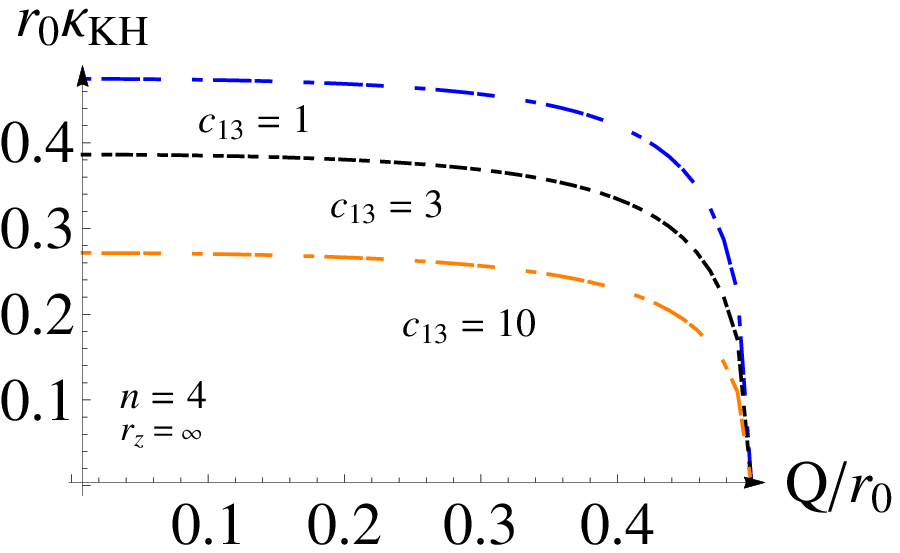}\includegraphics[width=6cm]{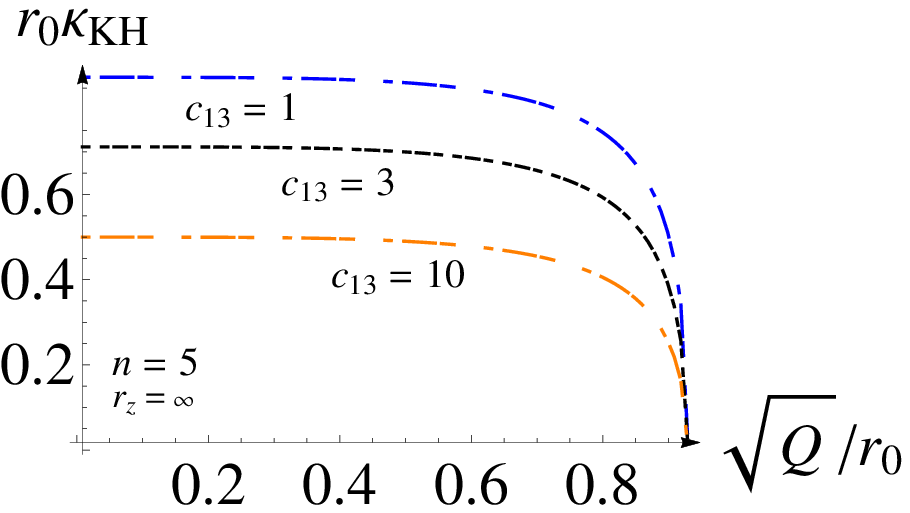}
\includegraphics[width=6cm]{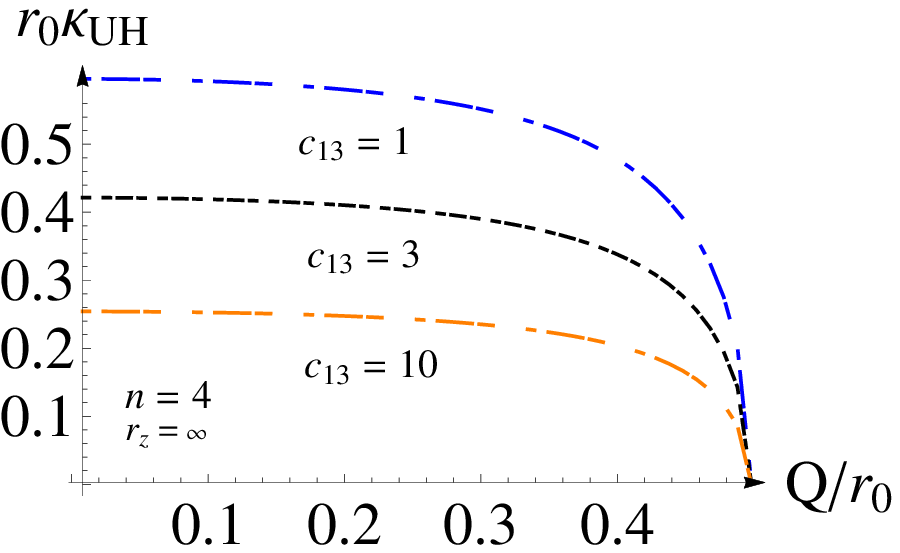}\includegraphics[width=6cm]{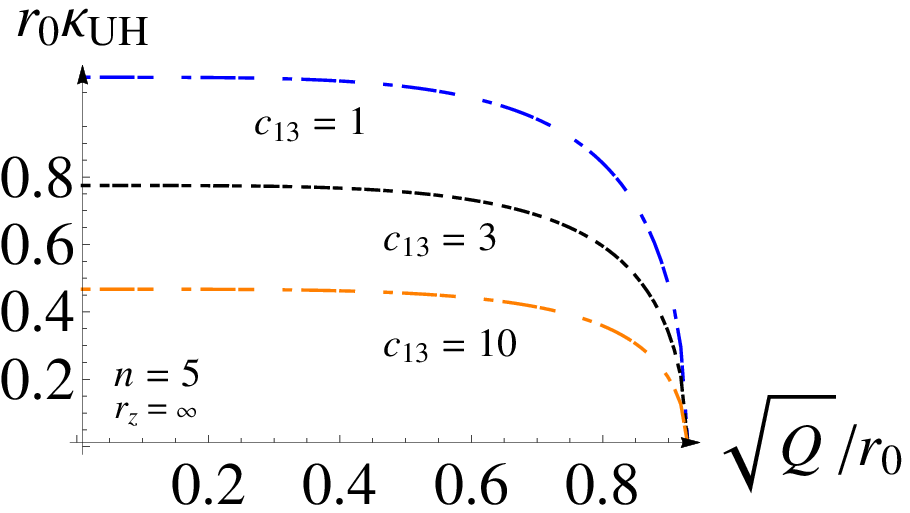}
\caption{The relation between $r_0\kappa_{KH}$, $r_0\kappa_{UH}$ and $Q^{1/(n-3)}/r_0$ in $4$ and $5$ dimensional asymptotically Anti-de Sitter spherical spacetimes with $c_{14}=0$, where we have chosen $\Lambda=-0.1/r_0^2$, $\alpha=1$ and $c_{123}=3$.} \label{figkNA}
\end{figure*}

\begin{figure*}[h]
\includegraphics[width=6cm]{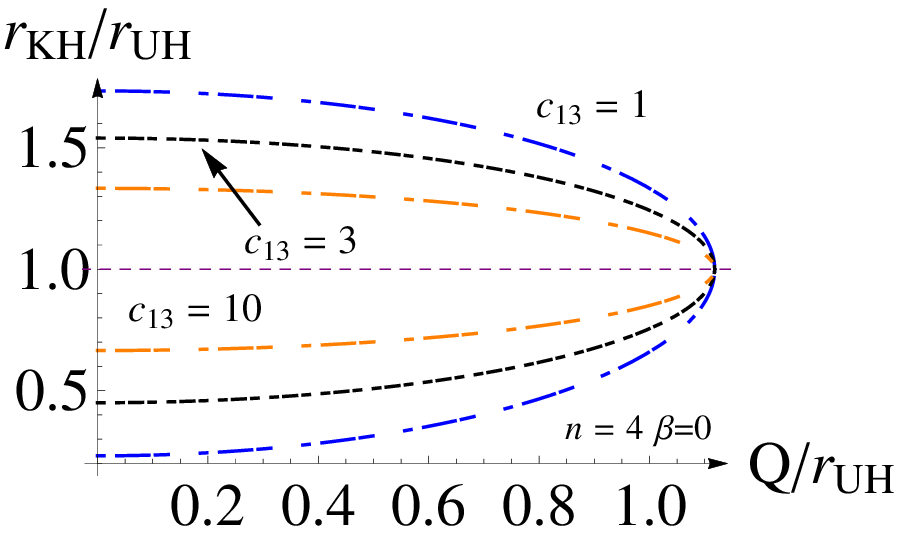}\includegraphics[width=6cm]{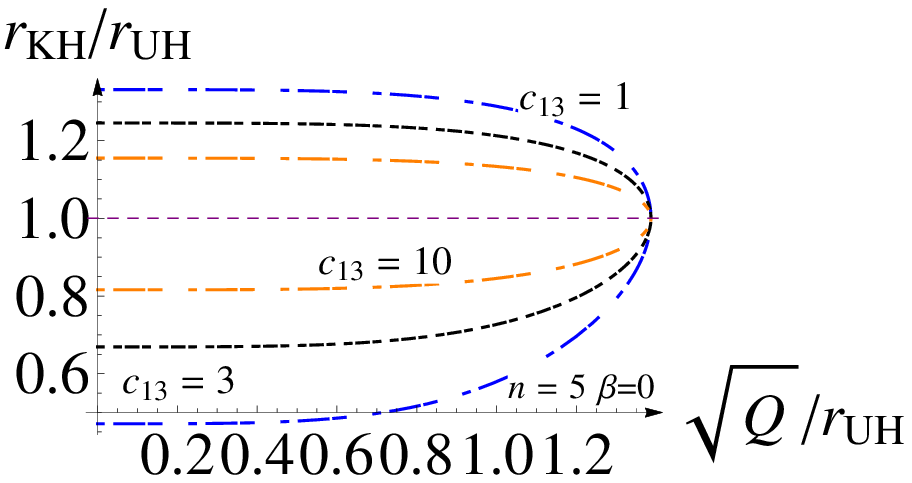}
\caption{The relation between $r_{KH}/r_{UH}$ and $Q^{1/(n-3)}/r_{UH}$ in $4$ and $5$ dimensional asymptotically Anti-de Sitter spherical spacetimes with $c_{123}=0$, where we have chosen $\Lambda=-0.1/r_0^2$, $\alpha=1$ and $c_{14}=0.3$.} \label{fighNB}
\end{figure*}

\begin{figure*}[h]
\includegraphics[width=6cm]{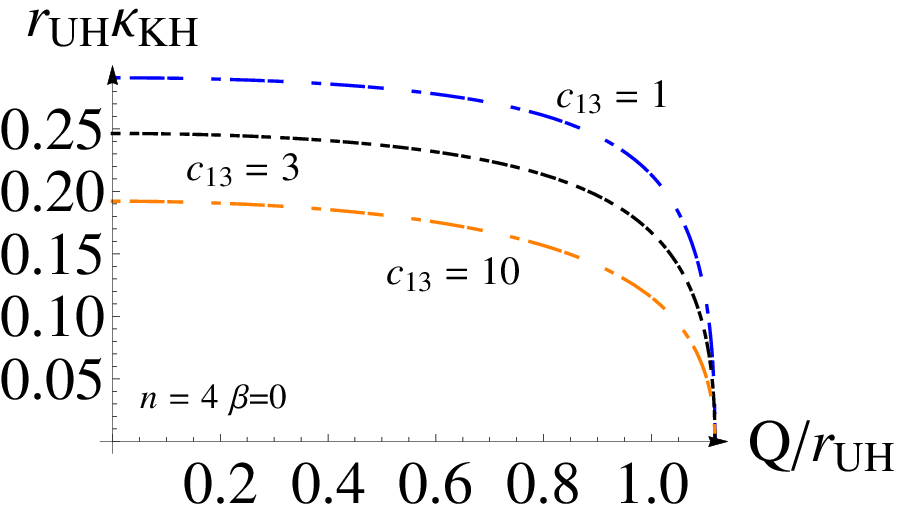}\includegraphics[width=6cm]{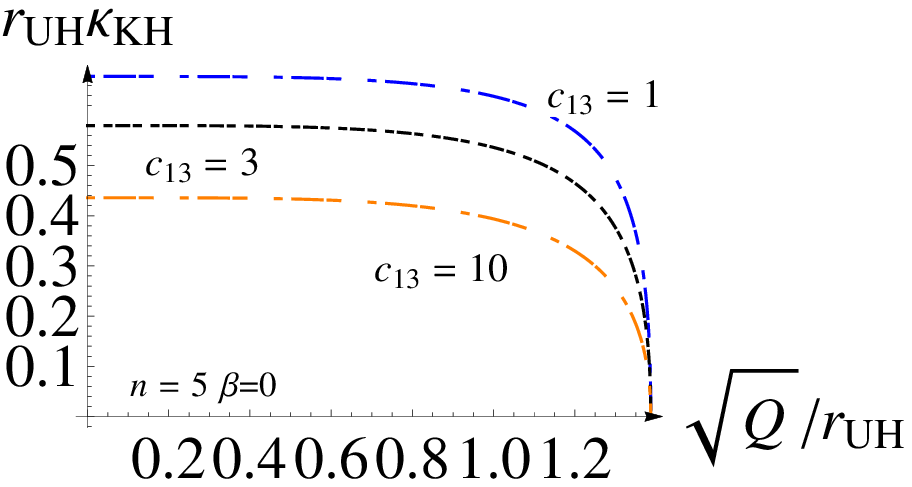}
\includegraphics[width=6cm]{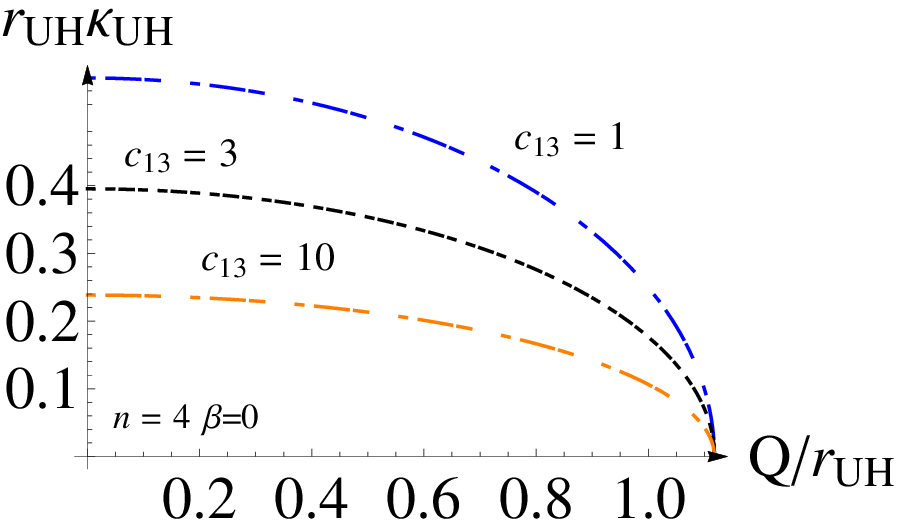}\includegraphics[width=6cm]{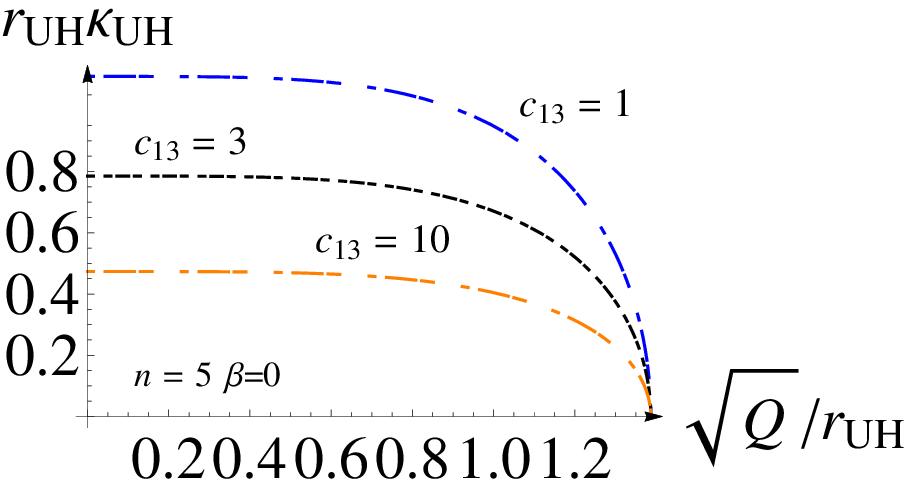}
\caption{The relation between $r_{UH}\kappa_{KH}$, $r_{UH}\kappa_{UH}$ and $Q^{1/(n-3)}/r_{UH}$ in $4$ and $5$ dimensional asymptotically Anti-de Sitter spherical spacetimes with $c_{123}=0$, where we have chosen $\Lambda=-0.1/r_0^2$, $\alpha=1$ and $c_{14}=0.3$.} \label{figkNB}
\end{figure*}

In asymptotically Anti-de Sitter spherical spacetimes, because of the same reason for de Sitter case, we don't need the conditions $r_z\rightarrow\infty$ at $c_{14}=0$ case and $\beta=0$ for the $c_{123}=0$ case.
We show the relation between horizon and surface gravity with $r_z\rightarrow\infty$ and $\beta=0$ in Figs.\ref{fighNA}-\ref{figkNB}.

The results for de Sitter and Anti-de Sitter spacetimes indicate that the properties of the black hole solution in those spacetimes are actually similar to those of asymmetrically flat case.
The outer and inner killing horizon merges with the universal horizon as the black hole solution transits to the extreme case, while the surface gravity vanishes.
However, it is interesting to note that the electric charge of the extreme case is different as $\beta\not=0$.

\subsection {asymptotically Anti-de Sitter planar spacetime \\$k=0$ and $\Lambda<0$}

Another well-known black hole solution is the Anti-de Sitter planar case, and it is very convenient to apply this spacetime to investigate holographical superconductor, which is an application of AdS/CFT Correspondence\cite{HHH1,HHH2,LA,LACW,LOA,LAW}.
In this section, we investigate the planar black hole of \ae ther-Einstein theory.

\begin{figure*}[h]
\includegraphics[width=6cm]{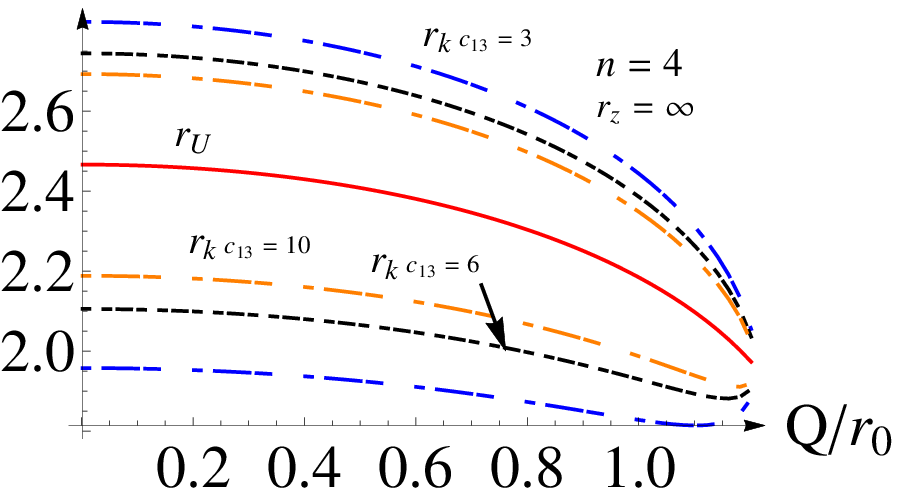}\includegraphics[width=6cm]{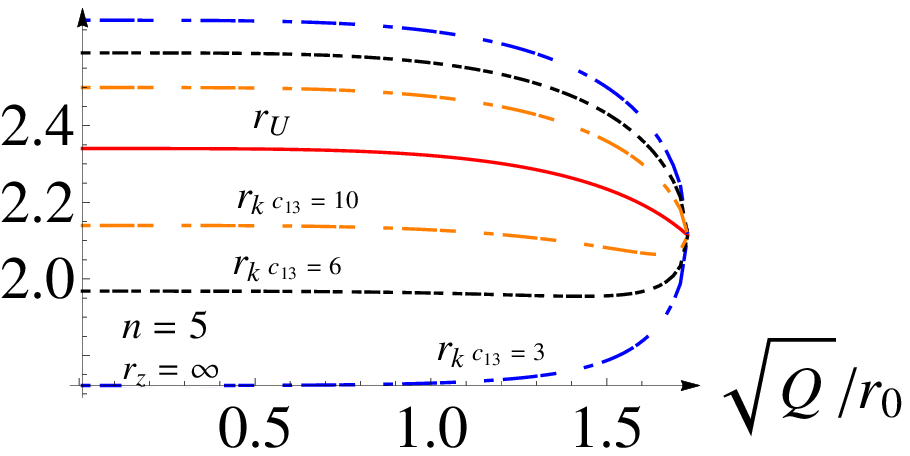}
\caption{The relation between $r_U\equiv r_{UH}/r_0$, $r_k\equiv r_{KH}/r_0$ and $Q^{1/(n-3)}/r_0$ in $4$ and $5$ dimensional asymptotically Anti-de Sitter planar spacetimes with $c_{14}=0$, where we have chosen $\Lambda=-0.1/r_0^2$, $\alpha=1$ and $c_{123}=3$.} \label{fighFA}
\end{figure*}

\begin{figure*}[h]
\includegraphics[width=6cm]{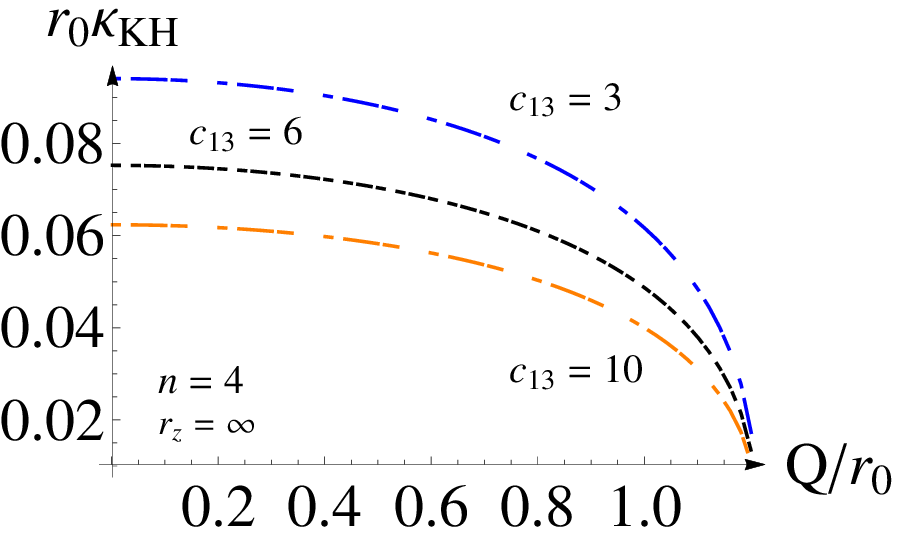}\includegraphics[width=6cm]{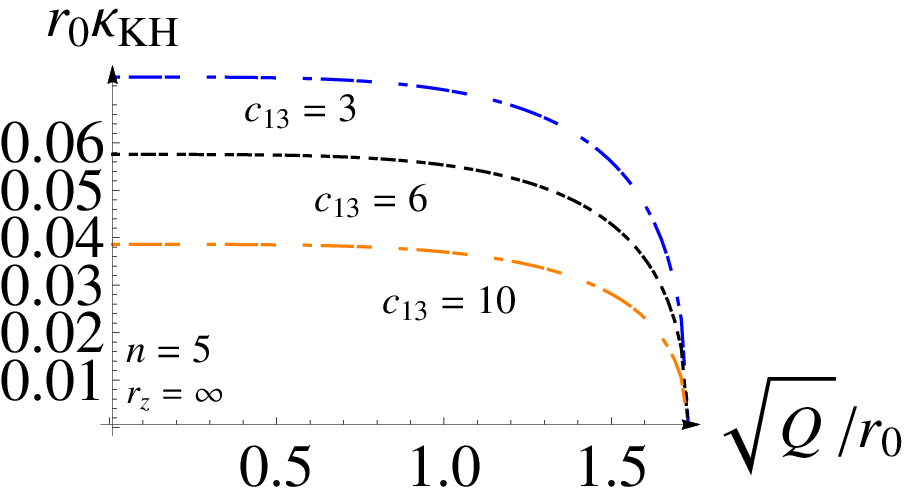}
\includegraphics[width=6cm]{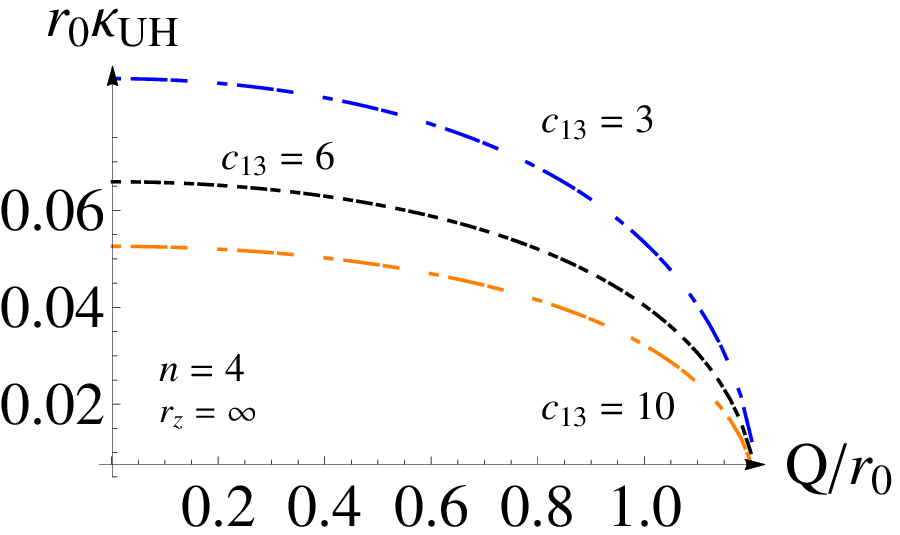}\includegraphics[width=6cm]{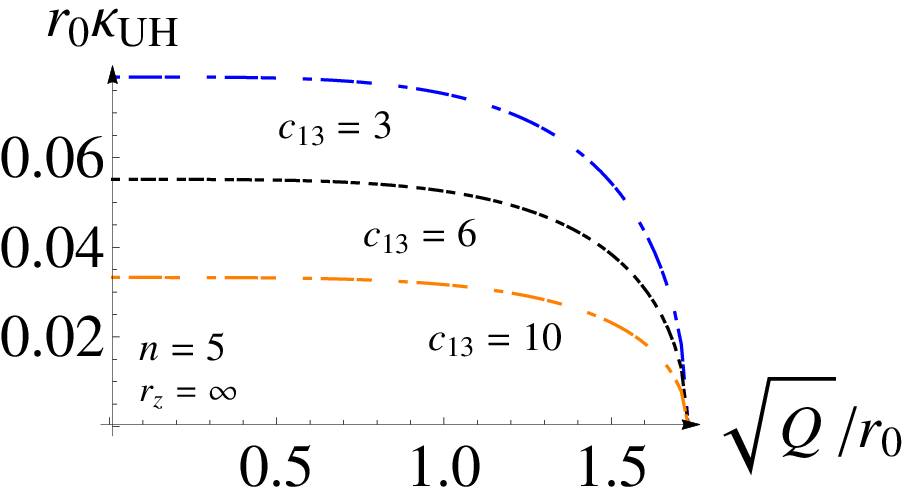}
\caption{The relation between $r_0\kappa_{KH}$, $r_0\kappa_{UH}$ and $Q^{1/(n-3)}/r_0$ in $4$ and $5$ dimensional asymptotically Anti-de Sitter planar spacetimes with $c_{14}=0$, where we have chosen $\Lambda=-0.1/r_0^2$, $\alpha=1$ and $c_{123}=3$.} \label{figkFA}
\end{figure*}

\begin{figure*}[h]
\includegraphics[width=6cm]{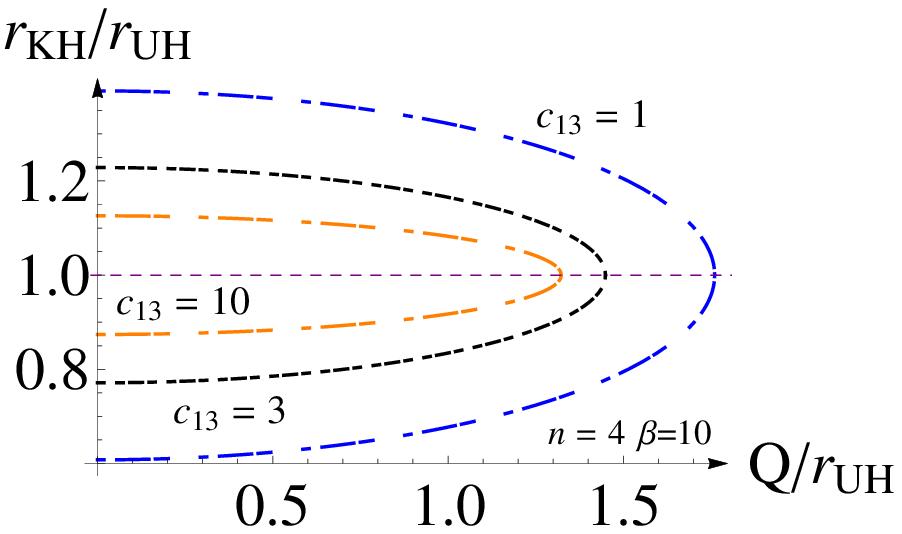}\includegraphics[width=6cm]{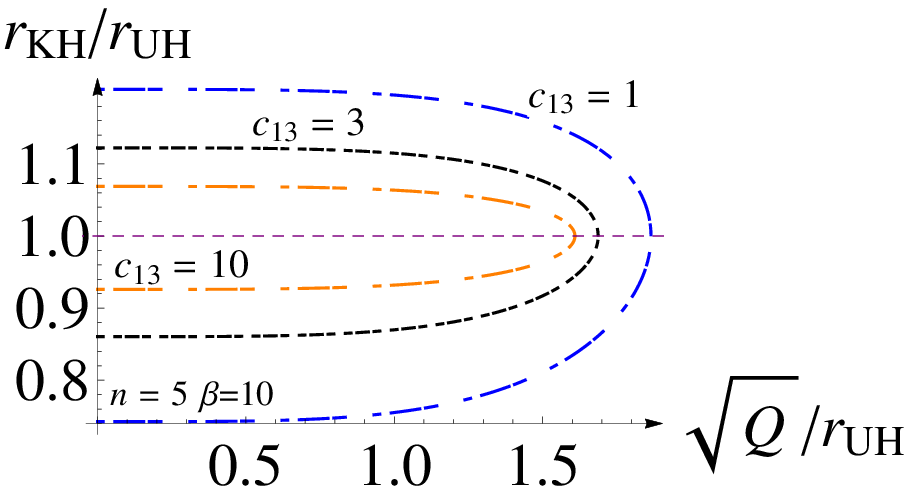}
\caption{The relation between $r_{KH}/r_{UH}$ and $Q^{1/(n-3)}/r_{UH}$ in $4$ and $5$ dimensional asymptotically Anti-de Sitter planar spacetimes with $c_{123}=0$, where we have chosen $\Lambda=-0.1/r_0^2$, $\alpha=1$ and $c_{14}=0.3$.} \label{fighFB}
\end{figure*}

\begin{figure*}[h]
\includegraphics[width=6cm]{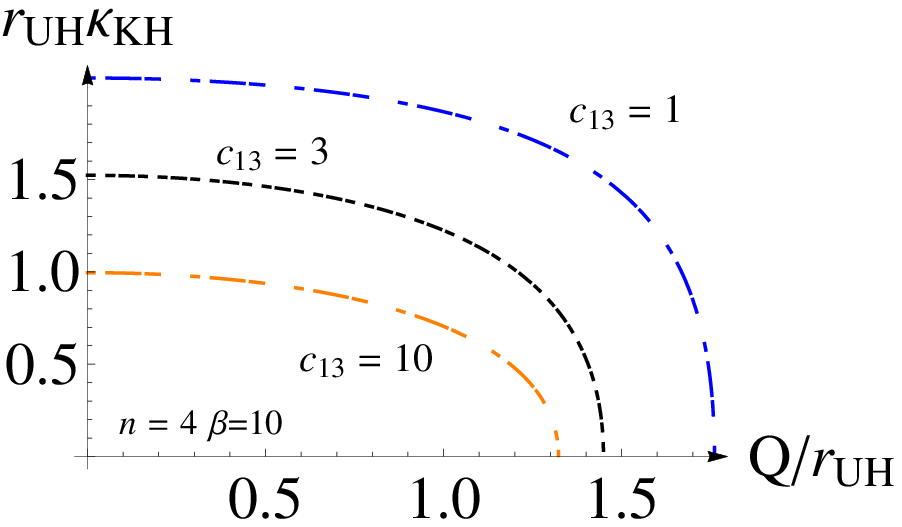}\includegraphics[width=6cm]{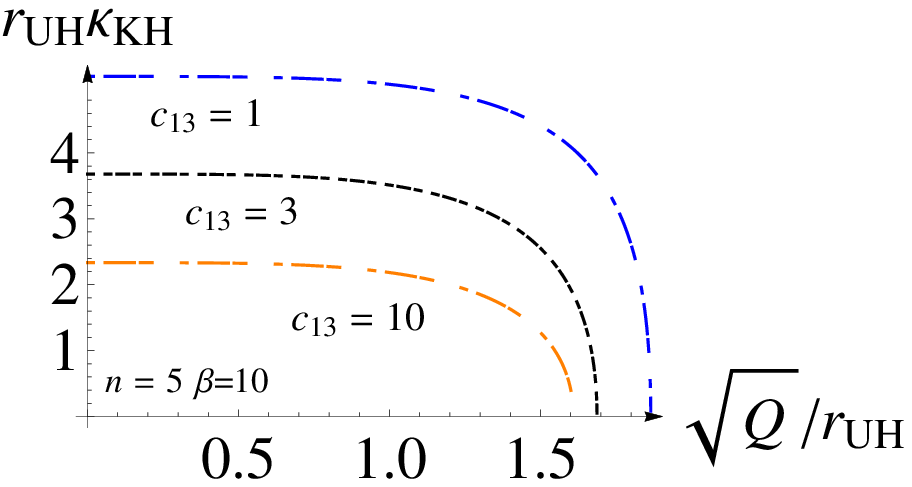}
\includegraphics[width=6cm]{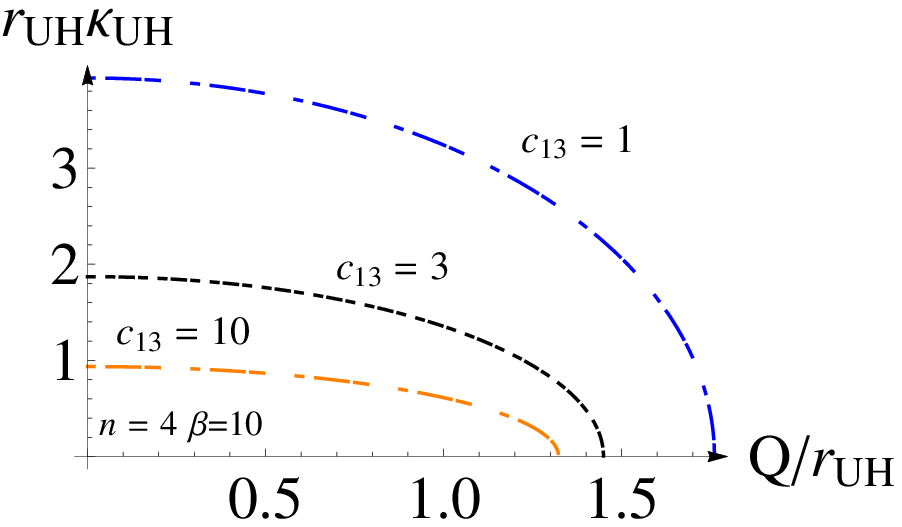}\includegraphics[width=6cm]{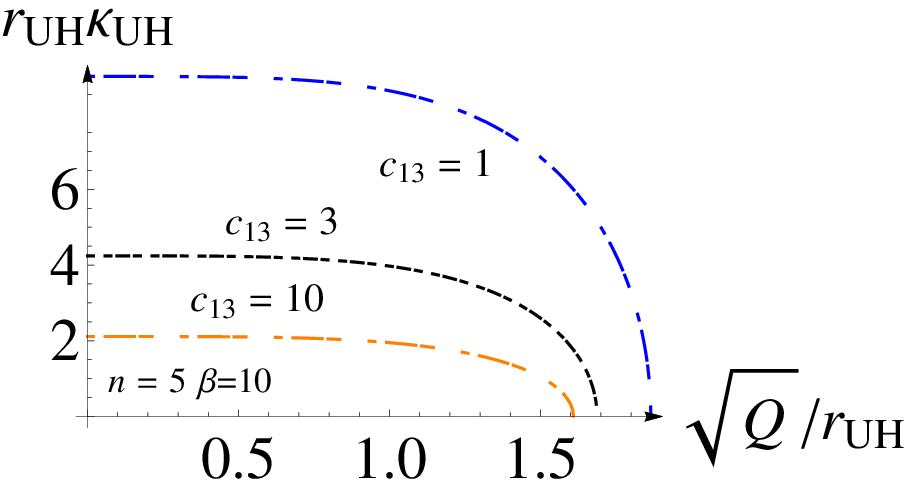}
\caption{The relation between $r_{UH}\kappa_{KH}$, $r_{UH}\kappa_{UH}$ and $Q^{1/(n-3)}/r_{UH}$ in $4$ and $5$ dimensional asymptotically Anti-de Sitter planar spacetimes with $c_{123}=0$, where we have chosen $\Lambda=-0.1/r_0^2$, $\alpha=1$ and $c_{14}=0.3$.} \label{figkFB}
\end{figure*}

In asymptotically Anti-de Sitter Planar spacetime, we show the relation between the horizon and surface gravity with $r_z\rightarrow\infty$ and $\beta=10$ in Figs.\ref{fighFA}-\ref{figkFB}.

\end{document}